\documentclass[aps,prb,twocolumn,floatfix,superscriptaddress,citeautoscript]{revtex4-2}

\usepackage{graphicx}
\usepackage{amsmath}
\usepackage{amssymb}
\usepackage{color, soul}

\makeatother

\newcommand{\beq}{\begin{equation}}
\newcommand{\eeq}{\end{equation}}
\newcommand{\beqy}{\begin{eqnarray}}
\newcommand{\eeqy}{\end{eqnarray}}
\newcommand{\beqs}{\begin{equation*}}
\newcommand{\eeqs}{\end{equation*}}
\newcommand{\bpm}{\begin{pmatrix}}
\newcommand{\epm}{\end{pmatrix}}

\newcommand{\vect}[1]{{\mathbf #1}}

\begin{document}

\title{Searching for the Kardar-Parisi-Zhang phase in microcavity polaritons}
\author{A. Ferrier}
\affiliation{ Department of Physics and Astronomy, University College London, Gower Street, London, WC1E 6BT, United Kingdom }
\author{A. Zamora}
\affiliation{ Department of Physics and Astronomy, University College London, Gower Street, London, WC1E 6BT, United Kingdom }
\author{G. Dagvadorj}
\affiliation{ Department of Physics and Astronomy, University College London, Gower Street, London, WC1E 6BT, United Kingdom }
\author{M. H. Szymańska}
\affiliation{ Department of Physics and Astronomy, University College London, Gower Street, London, WC1E 6BT, United Kingdom }

\begin{abstract}
Recent approximate analytical work has suggested that, at certain values of the external pump, the optical parametric oscillator (OPO) regime of microcavity polaritons may provide a long sought realisation of Kardar-Parisi-Zhang (KPZ) physics in 2D.  Here, by solving the full microscopic model numerically using the truncated Wigner method, we prove that this predicted KPZ phase for OPO is robust against the appearance of vortices or other effects.   For those pump strengths, spatial correlations in the direction perpendicular to the pump, and the distribution of phase fluctuations, match closely to the forms characteristic of the KPZ universality.  This strongly indicates the viability of observing KPZ behaviour in future polariton OPO experiments.  
\end{abstract}

\maketitle 

\section{Introduction}

The Kardar-Parisi-Zhang (KPZ) universality class offers a description of the long range behaviour in a wide variety of non-equilibrium systems.  Originally conceived as a model of growing surfaces \cite{PhysRevLett.56.889}, it has since been found to encompass a plethora of physical realisations, including growing cell colonies \cite{JPSJ.66.67,PhysRevE.82.031903}, burning paper \cite{PhysRevLett.79.1515,PhysRevE.64.036101,Eur.Phys.J.B.46.55}, and growing interfaces in liquid crystals \cite{Takeuchi2011,Takeuchi2012,PhysRevLett.124.060601}. Generally, however, with the exception of some progress with growing thin films \cite{EPL.105.50001,PhysRevB.89.045309} most experimental results have been limited to one dimension.  

One candidate for realising the KPZ universality in both 1D and 2D, is the phase dynamics of polariton condensates in semiconductor microcavities \cite{PhysRevX.5.011017,PhysRevB.91.045301,PhysRevB.92.155307,Rep.Prog.Phys.79.096001,PhysRevB.94.104520,PhysRevB.94.104521,keeling2016superfluidity,PhysRevLett.118.085301,PhysRevB.97.195453,PhysRevB.103.045302,Deligiannis_2021,fontaine2021observation}.  In the long range limit, only fluctuations of the free phase of the condensate remain relevant; eliminating all others leads to a KPZ equation for this phase.  Compared to the phase equation for thermal equilibrium condensates, the KPZ equation has additional non-linear terms arising from the drive and dissipation, which cause correlations to take a more rapidly decaying form \cite{PhysRevX.5.011017}.  In addition to potentially providing another much sought after experimental platform for investigating the 2D KPZ universality class, the phase being a compact variable offers a window into interesting new physics regarding the dynamics of vortices in the phase governed by the KPZ equation \cite{PhysRevB.94.104520,PhysRevB.94.104521,PhysRevLett.121.085704,PhysRevLett.125.265701,PhysRevLett.125.215301}.  

However, a consistent barrier has impeded reproducing these results in experiments, in that the length scales at which signs of KPZ are expected are unrealistically large compared to typical microcavities; this explains why previous numerical and experimental studies only observed behaviour analogous to equilibrium physics \cite{Caputo_2017,PhysRevX.5.041028}. Furthermore, with incoherent drive, free vortices were shown to proliferate beyond a length scale smaller than the KPZ length scale \cite{PhysRevB.94.104520}, rendering these polariton systems unsuitable for exploring the KPZ scaling phase. 

\begin{figure}[b]
\includegraphics[width=\columnwidth]{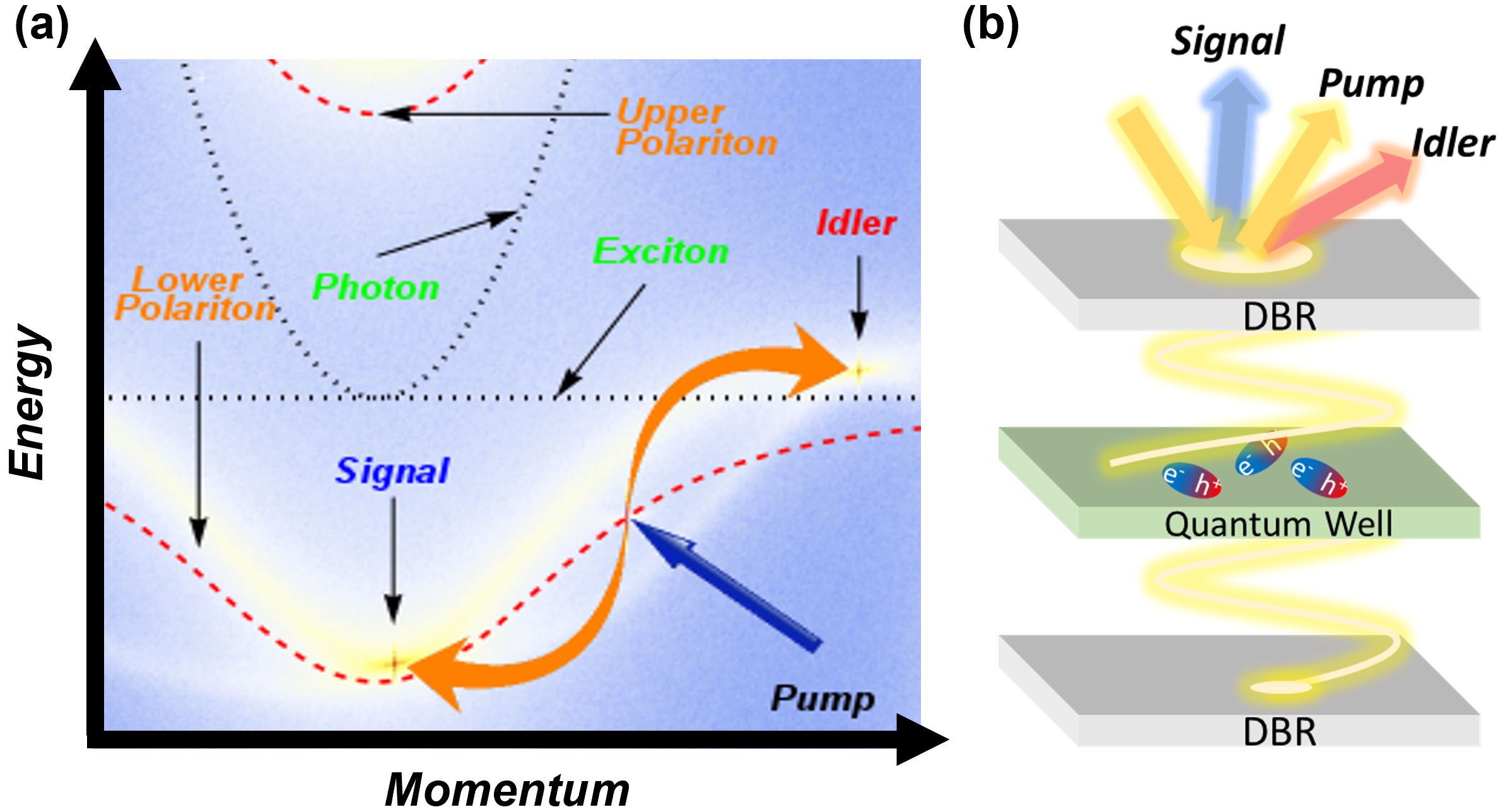}
\caption{Polaritons in semiconductor microcavities in the OPO regime.  (a) Typical spectrum of OPO showing signal, pump and idler modes on the lower polariton branch.  (b) Sketch of the system:  External laser drives the pump mode, which then scatters to occupy the signal and idler.  
\label{OPOdiag}}
\end{figure}

A potential way around this lies in an alternative regime of the microcavity polariton system, the optical parametric oscillator (OPO).  Here, rather than having a single condensate occupied by incoherent driving, a coherently driven ``pump" mode, scatters to occupy two other modes, the signal and the idler (see Fig.~\ref{OPOdiag}).  Recent work has shown \cite{PhysRevX.7.041006} that not only does the OPO regime map similarly to a KPZ equation for the free phase difference between the signal and idler, but it is also highly tunable by varying the coherent drive strength, even leading to a small window where KPZ behaviour should become observable at all length scales.

While that analytical study indicates a promising \mbox{direction} for the possibility of observing the 2D KPZ phase in polariton experiments, the question of whether this phase would also be destroyed by vortices, in analogy to the KPZ phase in incoherently driven systems, is still open.  Although discussed there, the analysis of \cite{PhysRevX.7.041006} cannot determine the behaviour of vortices, and whether a KPZ scaling phase without vortices or a vortex dominated phase due to altered vortex interactions \cite{PhysRevB.94.104520} should occur in the window predicted in \cite{PhysRevX.7.041006} ultimately remains ambiguous.  In addition, the analysis in \cite{PhysRevX.7.041006} contains a number of simplifications compared to real systems, which remain to be investigated to prove the viability of experimentally reproducing this behaviour.  Firstly, that a strictly three-mode (pump, signal, idler) model is considered, while in actual polariton OPO additional satellite states can also become relevant \cite{PhysRevB.98.165307, PhysRevB.71.115301}.  Secondly, that density fluctuations, which are neglected in the long range limit when mapping to the KPZ equation, may still be a relevant factor in real finite size systems or when satellite states are included, leading to, for example, spatially non-uniform condensates, pattern formation or time-dependent solutions.   

To address these questions, we now investigate the polariton OPO using full multimode stochastic simulations, not restricted by those approximations, to see if and how the signatures of the KPZ universality manifest when the pump strength is tuned to within the window found analytically.  Observing the predicted behaviour in this sort of numerical analysis should open the way to its replication in experiments, and hence the use of the OPO regime of polaritons in semiconductor microcavities as an experimental platform for exploring the KPZ universality in 2D.

\section{Model and Method}\label{section:TWA}

As illustrated by the spectrum in Fig.~\ref{OPOdiag} (a), strong coupling between the cavity photons and quantum well excitons in semiconductor microcavities leads to two branches of polaritons -- upper and lower.  Since the OPO regime is achieved by coherently driving the lower polariton branch, we neglect the upper polariton branch that will have negligible occupation, and consider a model with only lower polaritons.  The lower polaritons have a non-quadratic dispersion  
\mbox{$\omega_{lp}(\vect{k}) = \frac{1}{2}\left( \omega_c(\vect{k})+\omega_x -\sqrt{(\omega_c(\vect{k})-\omega_x)^2+\Omega_R^2} \right)$}, with $\omega_c(\vect{k})$ being the (quadratic) bare cavity photon dispersion, $\omega_x$ the exciton dispersion, which is approximately flat due its much larger mass, and $\Omega_R$ the Rabi frequency of exciton-photon coupling \cite{RevModPhys.85.299}.   

To study the system fully, we use stochastic simulations based on the truncated Wigner approximation (TWA) \cite{PhysRevB.72.125335,PhysRevB.79.165302,PhysRevLett.121.095302,PhysRevX.5.041028,PhysRevB.98.165307,PhysRevLett.125.095301}.  Unlike the three-mode model of OPO used for analytical calculations, our numerical method considers the full two dimensional multimode lower polariton field, which includes fluctuations in both density and phase, represented by a stochastic complex number field $\Psi\!\left(\vect{x},t\right)$, from which physical observables such as density and correlation functions can be calculated by appropriate averages over stochastic realisations.  By truncating the third order derivative terms, the equation for the evolution of the Wigner quasiprobability distribution can be reduced to the form of a Fokker-Planck equation. From this in turn, we can derive the following stochastic differential equation for trajectories of the stochastic complex number field $\Psi\!\left(\vect{x},t\right)$:
\begin{equation}
\frac{\partial\Psi\!\left(\vect{x},t\right)}{\partial t} =  -iH\Psi\!\left(\vect{x},t\right) +iF_p\!\left(\vect{x},t\right) + \sqrt{\frac{\kappa}{dV}}\Gamma\!\left(\vect{x},t\right) \, , \label{SGPE}
\end{equation}
with the differential operator $H$ defined as
\begin{equation*}
H = \omega_{lp}\!\left(-i\vect{\nabla}\right)-i\kappa+g\left(|\Psi\!\left(\vect{x},t\right)\!|^2-\frac{1}{dV}\right) \, ,
\end{equation*}
where $g$ is the polariton-polariton interaction strength, which we approximate as being momentum independent \cite{PhysRevB.98.165307}, $\kappa$ is the polariton decay rate, and $F_p\!\left(\vect{x},t\right) = f_pe^{i(k_px-\omega_pt)}$ is a coherent drive at momentum $k_p$ and frequency $\omega_p$.  $\Gamma\!\left(\vect{x},t\right)$ is a zero mean complex Wiener noise with $\langle\Gamma^*\!\left(\vect{x},t\right)\!\Gamma\!\left(\vect{x'},t'\right)\rangle = \delta_{\vect{x},\vect{x'}}\delta\!\left(t-t'\right)$.  Results of the TWA include all classical fluctuations and up to second order in quantum fluctuations \cite{PhysRevB.89.134310}, but discard higher order quantum effects, which only become relevant for much lower occupations or stronger interactions than considered here.  The area element $dV = a^2$ of the grid used to discretise space for numerical integration, where $a$ is the lattice spacing of this grid, plays a role in determining the validity of this approximation.  The TWA is appropriate under the condition that $\kappa\gg\frac{g}{dV}$.  

All physical quantities will be expressed in units derived from the parameters of the system: times in units of $2/\Omega_R$, lengths in units of $\sqrt{\hbar/(\Omega_Rm_c)}$, and energies in $\hbar\Omega_R/2$, where $m_c$ is the effective mass of photons in the cavity.  In these units, the other parameters are $g = 0.00118$, $\kappa = 0.045$, with the drive on resonance with the lower polariton dispersion at $k_p = 1.4$, $\omega_p = -0.42$.  We choose our energy scale, and the exciton-photon detuning, such that $\omega_x = \omega_c(0) = 0$.  These parameters are chosen to match those of modern polariton experiments with $\hbar\Omega_R = 4.4$meV and $m_c = 2.3 \times 10^{-5} m_e$ ($m_e$ the electron mass) \cite{sanvitto2010persistent}.  We consider a square area with side length $2L = 422.17544$ (roughly $366\mu$m, chosen to be on the order of real experimental microcavities), which is simulated on a $N\times N = 512\times 512$ point grid, giving $a = 0.8246$ ($dV = 0.6799$).  For these parameters, the upper and lower OPO thresholds occur at $f_p = 0.053$ and $f_p = 0.0135$ respectively \cite{PhysRevB.98.165307}.  

Physical observables are calculated within the TWA using the relation that averages over the Wigner distribution (i.e.~over stochastic realisations of our simulation) of products of the phase space variables correspond to quantum mechanical averages of the symmetrically ordered products of the relevant operators.  Of particular interest in this work is the first order spatial correlation of the signal mode
\beq
g^{(1)}_s\!\left(\vect{r}\right) = \frac{\langle\Psi^*_s\!\left(\vect{R}+\vect{r},t\right)\Psi_s\!\left(\vect{R},t\right)\rangle - \frac{\delta_{\vect{r}, \vect{0}}}{2dV}}
{\langle\Psi^*_s\!\left(\vect{R},t\right)\Psi_s\!\left(\vect{R},t\right)\rangle - \frac{1}{2dV}}\, , \label{g1def}
\eeq
where averages are taken over both stochastic realisations and the auxiliary position $\vect{R}$, and the signal field $\Psi_s\!\left(\vect{x},t\right)$ is isolated by filtering in momentum space (see Appendix \ref{appendix:Filter} for further details).

\section{Expected signatures of KPZ physics in polariton OPO}\label{section:KPZ}

In common theoretical descriptions, the polariton OPO system is approximated as consisting of three main modes: the pump mode, which is driven directly by an external laser, and the signal and idler modes, which become occupied by the parametric scattering of polaritons from the pump mode.  The phase of the pump mode is fixed by the external laser, but the system has one free phase, the relative phase $\theta$ between the signal and idler modes, resulting in a spontaneously broken U(1) symmetry with $\theta$ being the corresponding massless Goldstone mode \cite{PhysRevA.76.043807}.  In previous work \cite{PhysRevX.7.041006}, it was shown that the system of equations for the three modes reduces to an anisotropic KPZ equation \eqref{aKPZ} for the Goldstone mode $\theta$ in the long range limit: 
\beq
\partial_t \theta = D_x\partial_x^2\theta + D_y\partial_y^2\theta + \frac{\lambda_x}{2}\!\left(\partial_x\theta\right)^2 + \frac{\lambda_y}{2}\!\left(\partial_y\theta\right)^2 + \xi \, , \label{aKPZ}
\eeq
where $\xi\!\left(\vect r, t\right)$ is Gaussian noise with $\langle\xi\!\left(\vect r, t\right)\rangle = 0$ and $\langle\xi\!\left(\vect r, t\right)\xi\!\left(\vect{r'}, t'\right)\rangle = 2\Delta\delta\!\left(\vect r - \vect{r'}\right)\!\delta\!\left(t-t'\right)$.  For the OPO case, the diffusion coefficients $D_x$, $D_y$, non-linear coefficients $\lambda_x$, $\lambda_y$, and noise strength $\Delta$, depend in a non-trivial way on the physical parameters of the system (i.e.~$g$, $\kappa$, $f_p$, $k_p$, $\omega_p$).  The non-linear terms in particular arise from the breaking of thermal equilibrium conditions by the drive and dissipation.  

Without vortices, the KPZ equation leads to an algebraic decay of the spatial correlations of $\theta$, i.e.~$\langle\left(\theta\!\left(\vect R + \vect r, t\right)-\theta\!\left(\vect R, t\right)\right)^2\rangle \sim \tilde{r}^{2\chi}$, where $\tilde{r}$ is the distance rescaled to take into account the anisotropy, \mbox{$\tilde{r}^2 = (x/x_0)^2 + (y/y_0)^2$}, and $\chi \approx 0.39$ \cite{EPL.105.50001,PhysRevE.77.031134,PhysRevE.92.010101} is a universal critical exponent for 2D KPZ.  Under the assumption that we can neglect density fluctuations, these phase correlations would result in the spacial correlations of the momentum-filtered signal field $\Psi_s\!\left(\vect{x},t\right)$ showing a stretched exponential decay with distance 
\mbox{$g^{(1)}_s\!\left(\vect{r}\right) \sim e^{-\tilde{r}^{2\chi}}$} \cite{PhysRevX.7.041006}.  

\begin{figure}[h!]
\includegraphics[width=\columnwidth]{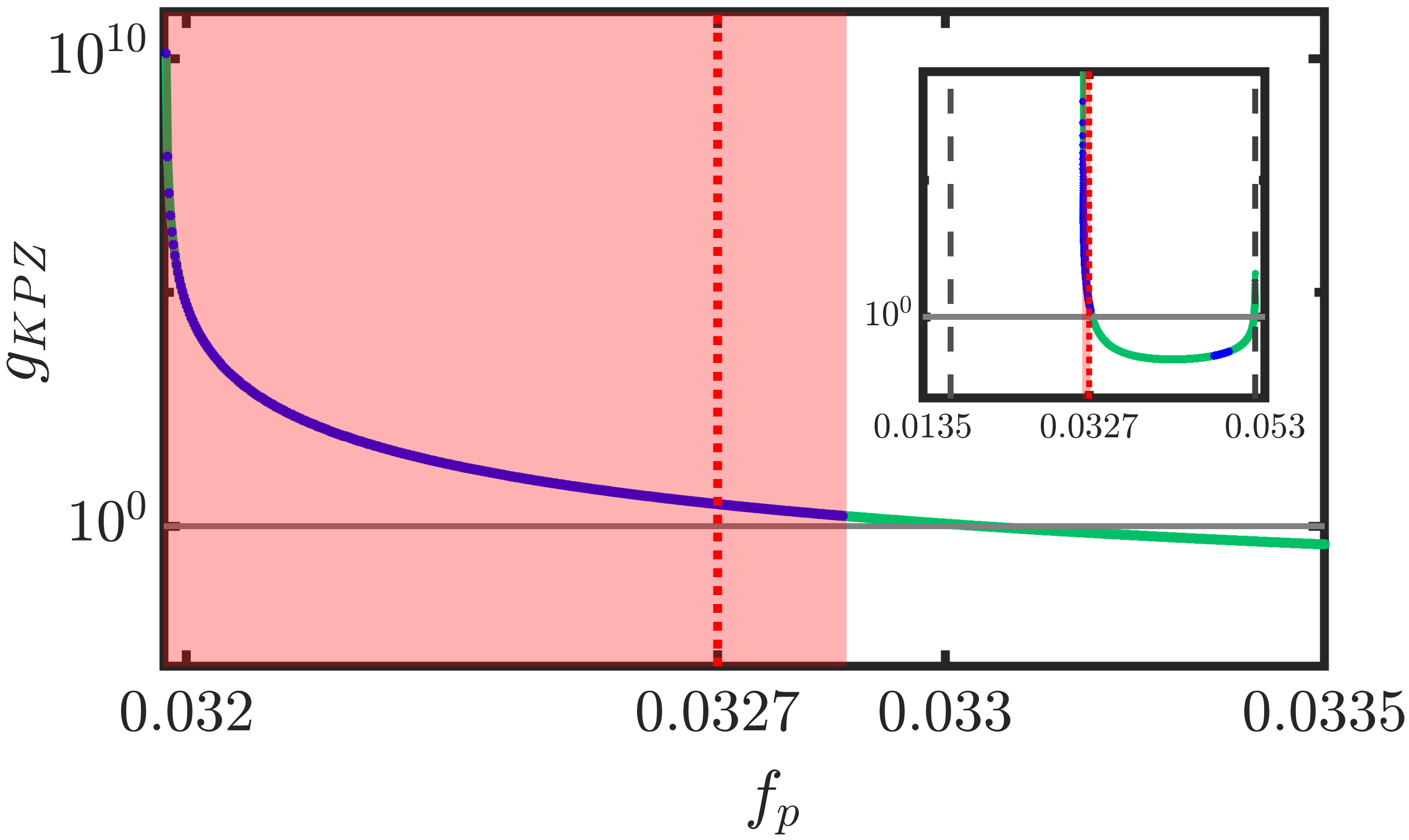}
\caption{KPZ non-linearity $g_{KPZ}$ as a function of the pump strength, $f_p$.  The blue line shows where the three-mode ansatz, assumed in deriving the KPZ equation for OPO, is stable. The window where we expect stretched exponential decay at all length scales is given by $g_{KPZ} \geq 1$ (horizontal line marks $g_{KPZ} = 1$).  The red shaded region shows where both three-mode OPO is stable and $g_{KPZ} \geq 1$.  The dotted vertical red line marks $f_p = 0.0327$, which we have chosen as representative  for full numerical analysis.  Parameters are $k_p = 1.4$, $k_s = 0.2084$, $\gamma = 0.045$, $\hbar\Omega_R = 4.4$meV, as in numerical simulations ($k_s$ is chosen to match that seen in the numerics at $f_p = 0.0327$).  Inset shows the full range of $f_p$ where OPO occurs, with the dashed vertical lines marking the OPO thresholds for our value of $k_s$.  
\label{fig_KPZnl}}
\end{figure}

In Ref.~\cite{PhysRevX.7.041006}, it was shown that within certain bounds in pump strength $f_p$, the non-linearity of the KPZ equation corresponding to the polariton OPO system can become large enough that the characteristic stretched exponential decay of spatial correlations should become observable at all length scales.  While an example of this range in $f_p$ was shown in the previous results \cite{PhysRevX.7.041006}, the analysis there ultimately depends on the exact value of the signal momentum, which in both numerical and real experiments is not an externally controlled parameter but chosen by the system as the OPO state forms, often in a way that is difficult to predict analytically \cite{PhysRevB.98.165307}.  As a result, we have first tested the behaviour at a selection of pump strengths around where we expect the window to be, and then checked where the analytical window is for the signal momentum that occurs in these numerical simulations of the full microscopic model at those pump strengths, and that the chosen $f_p$ actually falls within it.  Figure \ref{fig_KPZnl} shows the KPZ non-linearity $g_{KPZ} = \frac{\lambda_x^2\Delta}{D_x^2\sqrt{D_xD_y}}$ \cite{PhysRevX.7.041006} as a function of $f_p$ for the signal momentum found in the example cases.  The window where stretched exponential decay is expected to be easily visible is  where $g_{KPZ}$ is defined and $\geq 1$.  From this analysis, we choose the value $f_p = 0.0327$ to represent the behaviour within the KPZ window in our main results.

\section{Numerical results}\label{section:Compar}

\begin{figure}[b]
\includegraphics[width=\columnwidth]{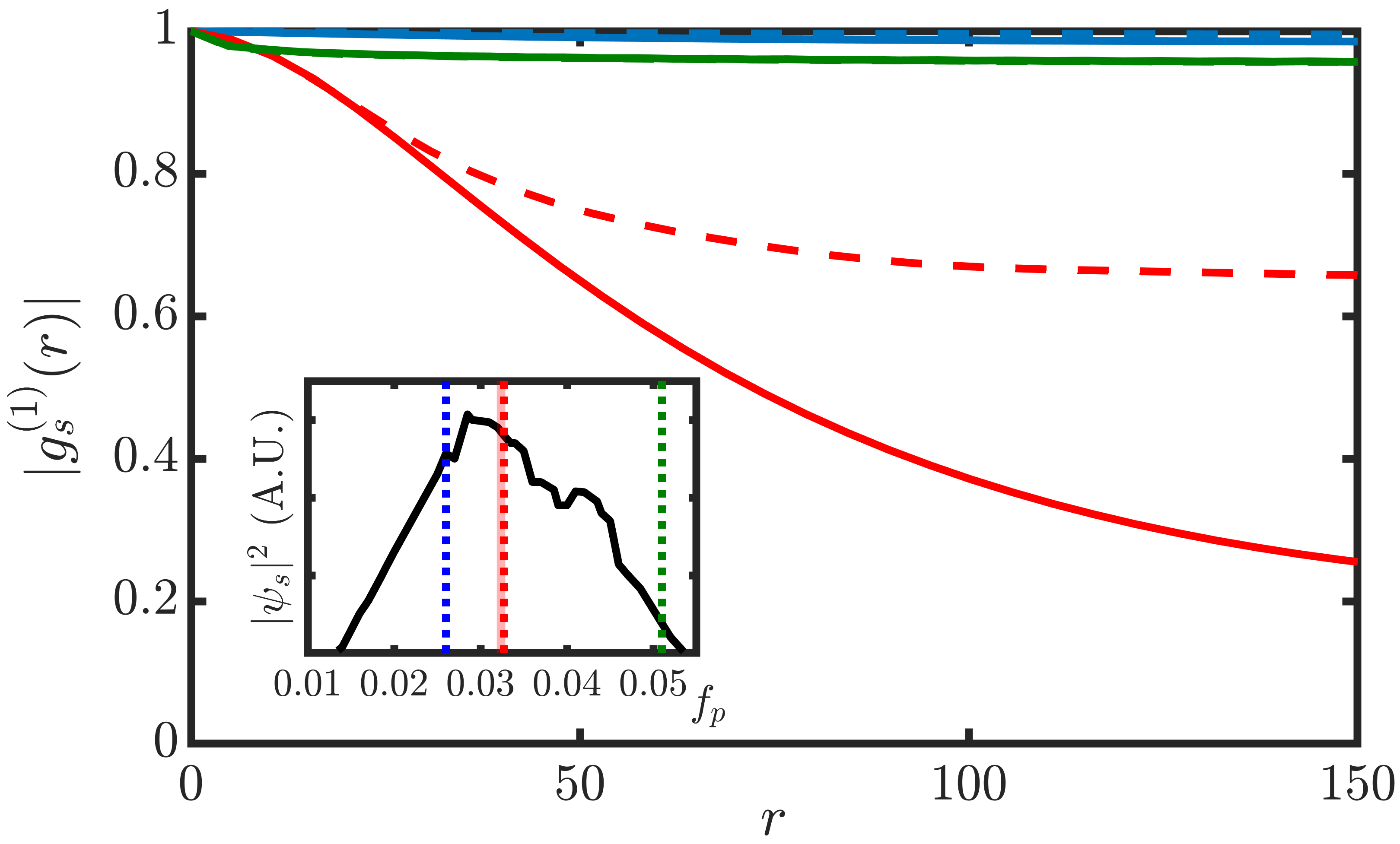}
\caption{Comparison of $g^{(1)}_s\!\left(\vect{r}\right)$ at different pump strengths $f_p$ for $\vect r$ taken along the $x$ (dashed lines) or $y$ (solid lines) directions.  For $f_p = 0.0327$ (red lines), within the KPZ window, correlations decay significantly faster, and with stronger anisotropy, than $f_p = 0.026$ (blue lines) and $f_p = 0.051$ (green lines) on either side of the window.  Inset:  Signal density versus pump strength $f_p$, vertical dotted lines indicate the chosen example values.  
\label{g1compar}}
\end{figure}

To begin with, we investigate how the behaviour within the KPZ window, which we choose to examine at $f_p = 0.0327$, differs from that outside it.  We run starting from a coherent mean field steady state until a steady state of the stochasitc simulations is reached for $g^{(1)}_s\!\left(\vect{r}\right)$ (see Appendix \ref{appendix:Conv} for details).  In Fig.~\ref{g1compar} we compare correlations along the $x$ and $y$ directions for three different values of $f_p$, one within the window, as mentioned at $f_p = 0.0327$, and one outside the window on either side, $f_p = 0.026$ in the middle of the OPO region and $f_p = 0.051$ near the upper threshold.  Values of the coefficients of equation \eqref{aKPZ} for each of these cases are given in Appendix \ref{appendix:Coef}.  

A number of distinct differences between the behaviour inside and outside the KPZ window are already recognisable just from inspection.  Firstly, within the window $g^{(1)}_s\!\left(\vect{r}\right)$ decays much faster with distance in both directions.  This is consistent with signatures of KPZ becoming observable within this window since, far from the Berezinskii–Kosterlitz–Thouless (BKT) transition, without the occurence of the KPZ regime, there would only be a very slow algebraic decay of correlations in the quasi-ordered state, as indicated by the blue and green lines in Fig.~\ref{g1compar}.  This dramatic change should make it clear when the KPZ regime is reached when sweeping the driving strength in experiments.  Note that the quasi-condensate density of the signal at the pump strength chosen inside the KPZ window (marked by red dotted vertical line in Fig.~\ref{g1compar}) is significantly larger than at the considered pump strengths outside of the KPZ window (marked by the blue and especially by the green dotted vertical line in Fig.~\ref{g1compar}).  In the usual case of algebraically decaying correlations associated with the quasi-ordered state in two dimensions \cite{PhysRevX.5.041028}, lower densities always mean faster decay of coherence.  The fact that we observe a much faster decay of correlations for a case with significantly larger signal density than the other cases, indicates strongly that the physics is dominated by a different type of fluctuations than in equilibrium 2D quasi-condensates.  Curiously, the KPZ window also displays significant anisotropy in the behaviour of $g^{(1)}_s\!\left(\vect{r}\right)$, that is not observable in the almost constant correlations outside it; for the example inside the KPZ window, correlations decay much faster with distance in the $y$ direction than in $x$ (compare the red lines in Fig.~\ref{g1compar}).

\section{Fitting the form of spatial correlations}\label{section:Fitting}

With a clear indication that something significant is occurring within the KPZ window, we now investigate how well this behaviour fits to the stretched exponential form predicted by the KPZ equation.  For this purpose we fit the form of $g^{(1)}_s\!\left(\vect{r}\right)$ to three different models: algebraic decay ($g^{(1)}_s\!\left(r\right) \sim r^{-\alpha}$), exponential decay ($g^{(1)}_s\!\left(r\right) \sim e^{-\frac{r}{r_0}}$), and stretched exponential decay (as defined above) with the universal exponent $\chi = 0.39$.  

\begin{figure}[h!]
\includegraphics[width=\columnwidth]{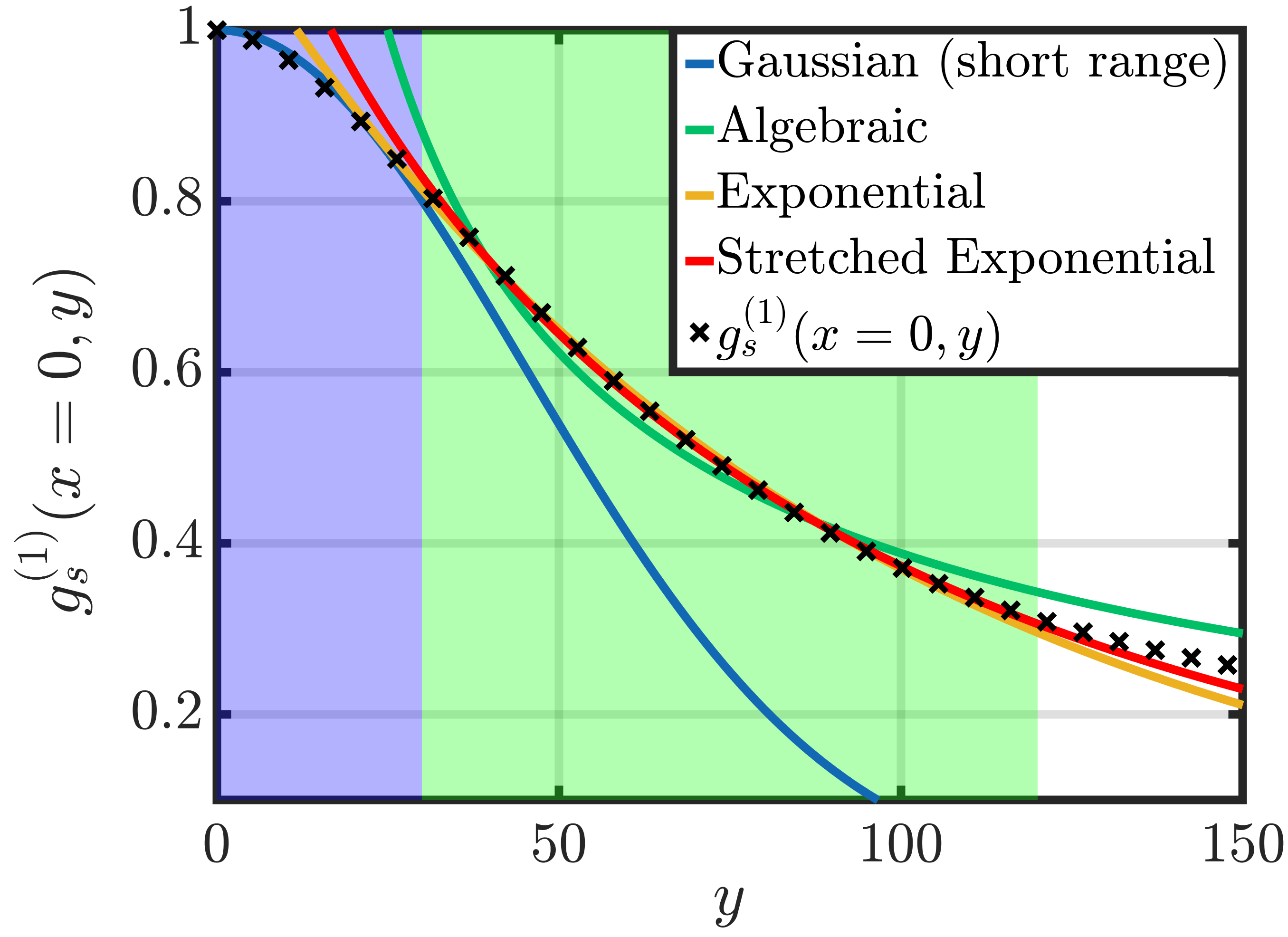}
\caption{$g^{(1)}_s\!\left(x = 0, y\right)$ showing fits to Gaussian form (blue line) of short range correlations at $y<30$ (blue shaded region), and fits to algebraic (green line), exponential (yellow line), and stretched exponential (red line) decay in the (green shaded) region $30 < y < 120$.  
\label{g1Allfits327}}
\end{figure}

For the case $f_p = 0.0327$, correlations in the $y$ direction fit well to the stretched exponential decay with $\chi = 0.39$.  In Appendix \ref{appendix:Exp}, we also determine the value of $\chi$ from our data, and confirm it agrees with the expected 
$\chi \approx 0.39$, finding $\chi = 0.41(3)$ when left as a free parameter in fitting $g^{(1)}_s\!\left(x = 0, y\right)$, and $\chi=0.38(3)$ from a power law fit of the corresponding phase correlations.  In \mbox{Fig.~\ref{g1Allfits327}} we show the algebraic, exponential, and stretched exponential fits to $g^{(1)}_s\!\left(x = 0, y\right)$ for $f_p = 0.0327$.  For the fits we exclude both the short range behaviour, indicated by the blue region in Fig.~\ref{g1Allfits327}, which is not expected to obey the stretched exponential form that is found in the long-range limit, and the furthest points that are most significantly affected by the finite size of the system and periodic boundary conditions.  Further discussion of the exact justification of our fitting bounds is included in Appendix \ref{appendix:Bounds}.  For this central portion of the correlation function, as measured by the coefficient of determination $R^2$, the stretched exponential decay ($R^2 = 0.9995$) fits significantly better than the algebraic decay ($R^2 = 0.9784$), which would be expected outside the KPZ regime, and slightly better than the pure exponential decay ($R^2 = 0.9986$).  Additionally, we do not expect a pure exponential decay of correlations since this is associated with strong disorder, i.e.~for our type of system, the presence of vortices, and we do not find any vortices in the momentum-filtered signal field $\Psi_s\!\left(\vect{x},t\right)$.  As indicated earlier, our case is far away from the BKT transition and deep in the quasi-condensate phase, where without the KPZ type fluctuations we would expect the usual slow algebraic decay of correlations.  The correlations in the $x$ direction do not match as well to the stretched exponential decay (see Fig.~\ref{g1sxfits}).  We believe this results from the effect of additional satellite modes (arranged along $k_x$), which are not accounted for in the analytical mapping from polariton OPO to the KPZ equation, and may lead to additional effects (see Appendix \ref{appendix:Xdir} for details).

\section{Distribution of phase fluctuations}\label{section:Pfluc}

\begin{figure}[h!]
\includegraphics[width=\columnwidth]{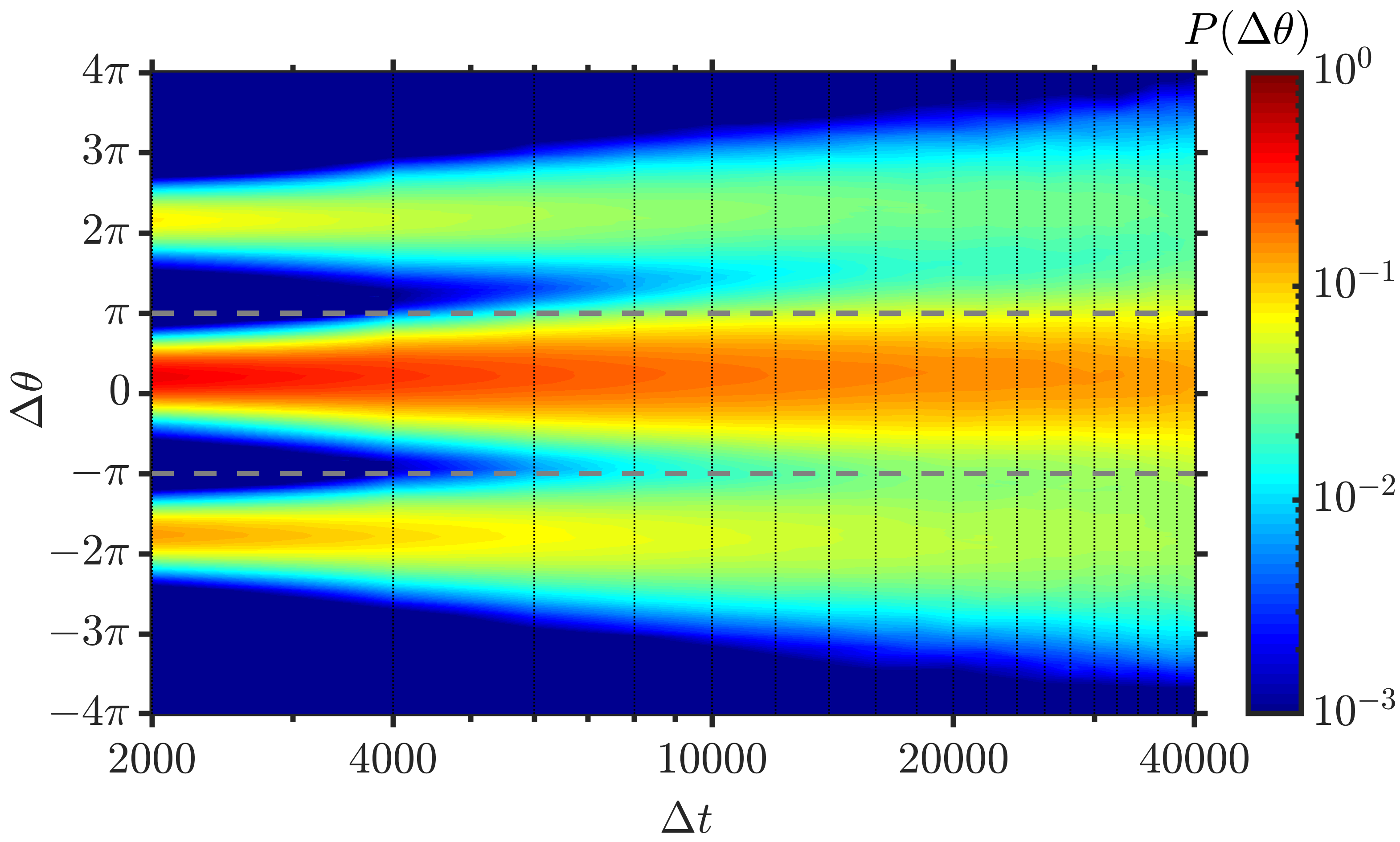}
\caption{Variation of distribution $P(\Delta\theta)$ of phase fluctuations of filtered signal mode $\Delta\theta$ with the time interval $\Delta t$ from reference time $t_0 = 120000$.  Vertical lines mark the time samples, while the horizontal dashed grey lines mark out the boundaries $\Delta\theta = \pm\pi$.  Colormap is truncated below $10^{-3}$.  
\label{PdPh_Growth}}
\end{figure}

To further elucidate the connection between the behaviour of the signal mode and the physics of the KPZ universality, we investigate the distribution of fluctuations in the phase of the signal mode.  The model-independent forms of such distributions displayed by other systems within the 2D KPZ universality under specific conditions have been established in the literature \cite{PhysRevLett.109.170602,PhysRevE.88.042118}.  We follow a similar pattern of analysis as has been used in recent work on the 1D KPZ universality in 1D incoherently driven exciton-polariton systems \cite{PhysRevB.97.195453,Deligiannis_2021,fontaine2021observation}.  Starting from a reference time in the steady state, $t_0 = 120000$, we measure the phase difference from time $t_0$ to $t_0 +\Delta t$,
$\delta\theta(\vect{x}, t_0, \Delta t) = \theta(\vect{x}, t_0+\Delta t) - \theta(\vect{x}, t_0)$, at each point $\vect{x}$.  The phase difference $-\pi<\delta\theta\leq\pi$ is unwound in time so that it may take unbounded values as it evolves.  The unwound phase difference is then expected to evolve according to
\beq
\delta\theta(\vect{x}, t_0, \Delta t)  \sim \omega_0\Delta t + (\Gamma\Delta t)^\beta \tilde{q}\, , \label{DthetaGrow}
\eeq
where $\beta$ is the universal growth exponent, $\tilde{q}$ is a random variable, and $\omega_0$, $\Gamma$ parameterise the growth of the mean and variance of $\delta\theta(\vect{x}, t_0, \Delta t)$, respectively.  We remove the deterministic part of the evolution of $\delta\theta$ to define a new, zero mean fluctuation, \mbox{$\Delta\theta(\vect{x}, t_0, \Delta t) = \delta\theta(\vect{x}, t_0, \Delta t) - \langle\delta\theta(\vect{x}, t_0, \Delta t)\rangle$}, where the average is over both realisations and position 
$\vect{x}$.  We take each point in space $\vect{x}$ for each stochastic realisation of our simulations as a separate sample, for a total of $512\times512\times400$ samples, to generate the distribution of the fluctuations $P(\Delta\theta)$ at each time sample 
$\Delta t$.  The evolution of this distribution is shown in Fig.~\ref{PdPh_Growth}.  

\begin{figure}[h!]
\includegraphics[width=\columnwidth]{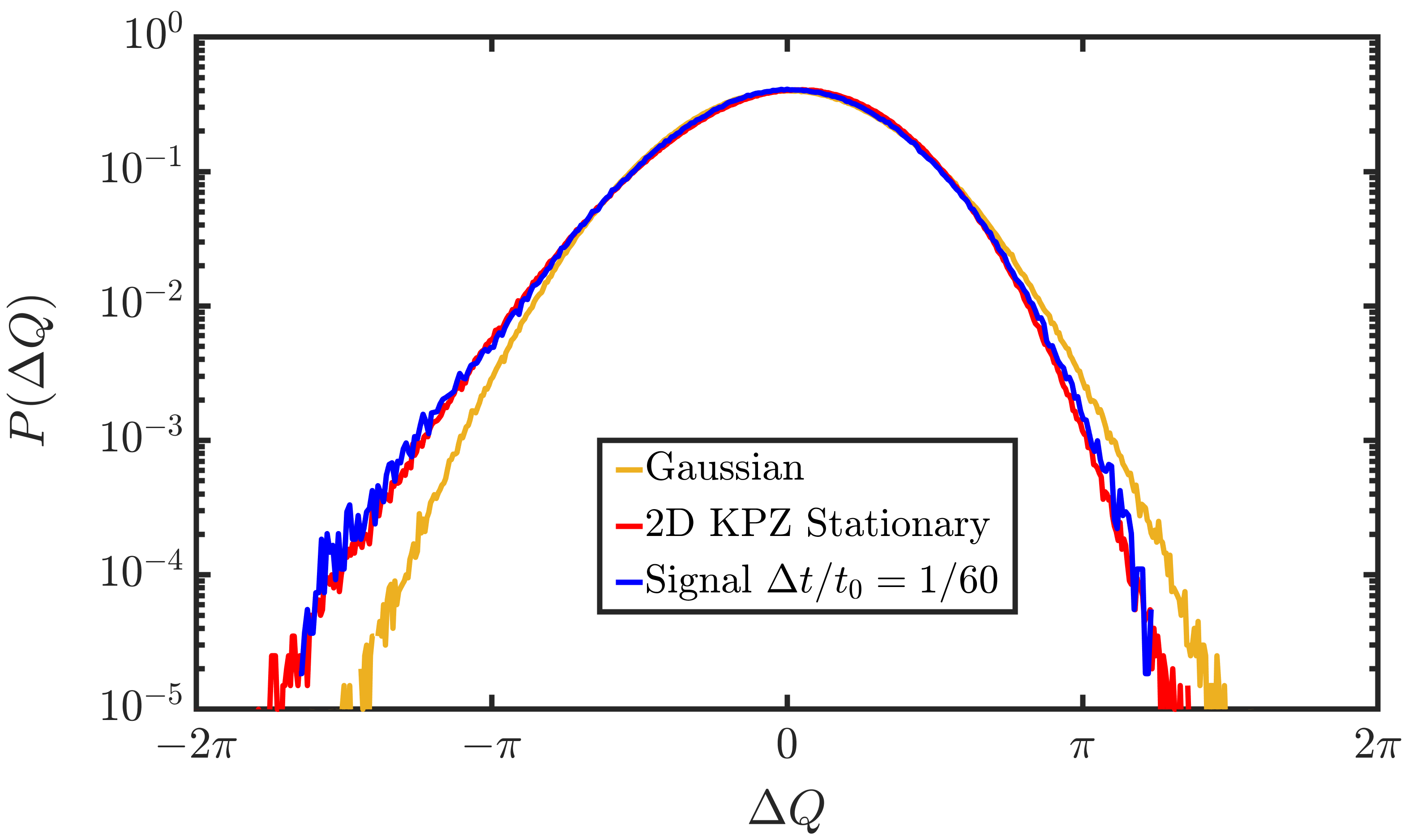}
\caption{Distribution $P(\Delta Q)$ of rescaled phase fluctuations of filtered signal mode $\Delta Q$ at $\Delta t/t_0 = 1/60$ (blue) compared with negative skewness version of the universal stationary distribution for 2D KPZ (red) and Gaussian distribution (yellow), each generated from $10^7$ samples of random variables from the corresponding distribution.  
\label{PdQ}}
\end{figure}

Similar to as was seen in the 1D case \cite{fontaine2021observation}, the distribution shows multiple lobes separated by $\pm2\pi$, which occur due to the unwinding of the phase difference into a continuous variable.  The width of each lobe can be seen to grow with time, as suggested by the form \eqref{DthetaGrow}.  Unfortunately, the multiple lobes begin to overlap each other after the first few time samples, making it difficult to use this time evolution to extract the value of the critical exponent $\beta$.  One way we can make use of this data, however, is to compare a single lobe of the distribution to the universal limit distributions for 2D KPZ \cite{PhysRevLett.109.170602,PhysRevE.88.042118}.  In particular, in the limit $\Delta t \ll t_0$, we expect that a single lobe of our distribution $P(\Delta\theta)$ should take the form of the stationary distribution associated with the 2D KPZ universality, as was determined in Ref.~\cite{PhysRevE.88.042118}.  To check this, we examine the central lobe of the distribution at the earliest time sample 
$\Delta t = 2000$, by keeping only the samples within the range $-\pi<\Delta\theta\leq\pi$; we then rescale this section of the distribution by defining a new variable,
\beq
\Delta Q = \frac{\Delta\theta - \langle\Delta\theta\rangle}{\sqrt{\langle\Delta\theta^2\rangle-\langle\Delta\theta\rangle^2}}\, , 
\eeq
which by definition has zero mean and unit variance over the range \mbox{$-\pi<\Delta\theta\leq\pi$}.  In Fig.~\ref{PdQ}, the new distribution $P(\Delta Q)$ is then compared to a zero mean, unit variance Pearson distribution with skewness \mbox{$s = \langle\Delta Q^3\rangle/\langle\Delta Q^2\rangle^{\frac{3}{2}}=-0.244$} and excess kurtosis $k = \langle\Delta Q^4\rangle/\langle\Delta Q^2\rangle^2 - 3 = 0.177$, corresponding to a negative skewness version of the stationary distribution of 2D KPZ \cite{PhysRevE.88.042118}.  We can see that the distribution $P(\Delta Q)$ generated from our data and rescaled to have unit variance (blue curve in Fig. \ref{PdQ}) matches closely to the universal stationary distribution of 2D KPZ (red curve in Fig. \ref{PdQ}), and is distinguishably different from a simple Gaussian distribution (yellow curve in Fig. \ref{PdQ}).  This adds further weight to the claim that our results arise from the connection of the polariton OPO to the 2D KPZ universality.  

\section{Decay of vortices}\label{section:Vort}

\begin{figure}[h!]
\includegraphics[width=\columnwidth]{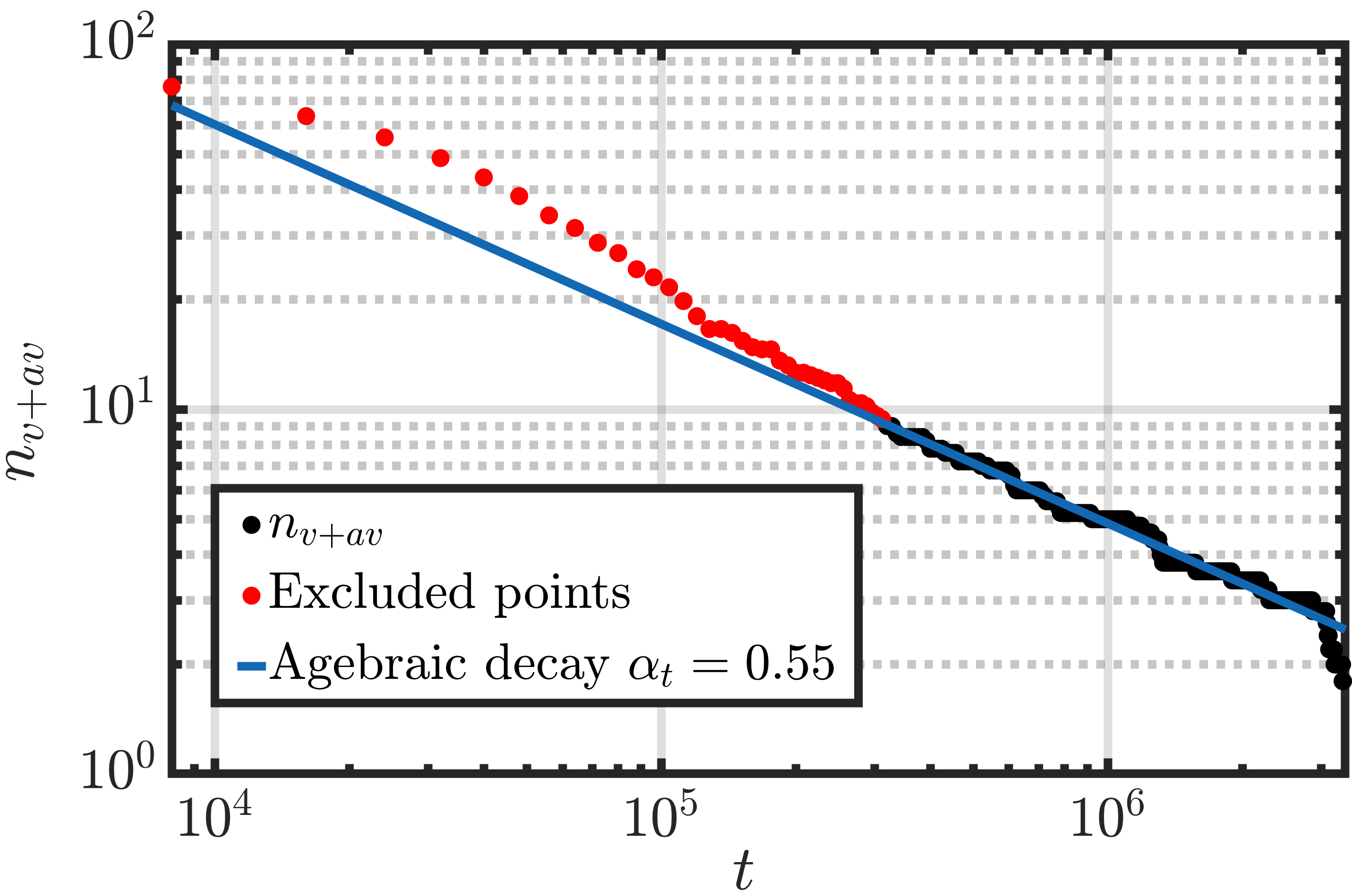}
\caption{Average number of vortices and antivortices $n_{v+av}$ with time, when starting from highly disordered initial conditions with $f_p = 0.0327$.  Blue line shows fit to algebraic decay of $n_{v+av}$ with time at late times $t \geq 320000$, with exponent 
$\alpha_t = 0.55$.  \label{vortexdecay}}
\end{figure}

To confirm that this phase without vortices is the true steady state, we also investigate how the system evolves for long times starting from highly disordered initial conditions, the opposite case to the completely coherent initial conditions used above.  We run 10 more realisations at $f_p = 0.0327$ in this way, and measure the decay of the average number of vortices and antivortices $n_{v+av}$ with time, shown in Fig.~\ref{vortexdecay}, to see if it continues to decay towards the vortex free steady state.  At late times $t \geq 320000$, $n_{v+av}$ is fitted to an algebraic decay in time, $n_{v+av} = \left(t/t_0\right)^{-\alpha_t}$, with $\alpha_t =0.55$, eventually falling below one pair on average.  However, the decay of vortices is notably slower than the usual equilibrium-like phase ordering, $\alpha_t = 1$ (with a logarithmic correction), previously seen in polariton OPO simulations \cite{PhysRevLett.121.095302}, which could be indicative of the altered interactions of vortices under the KPZ equation predicted in \cite{PhysRevB.94.104520}.  

\section{Summary and Outlook}

We have shown that evidence of KPZ in polariton OPO can be seen clearly in numerical solutions of the system's microscopic equations of motion.  Although we do observe additional complexity compared to the three-mode analytical model used to predict this behaviour \cite{PhysRevX.7.041006}, the distinctive stretched exponential decay of first order spatial correlations is still clearly visible in the direction perpendicular to the pump momentum.  In this direction, the roughness exponent $\chi$ characteristic of 2D KPZ behaviour can be found from both this and the corresponding algebraic scaling of phase correlations.  Furthermore, the distribution of the signal's phase fluctuations also match the universal form expected in the stationary limit for the 2D KPZ class.  We also prove using different initial conditions that the steady state of the system does not contain any vortices, in stark contrast to what has been seen in the isotropic compact KPZ equation \cite{PhysRevLett.125.265701} and expected in incoherently driven systems \cite{PhysRevB.94.104520}.  The dramatic change in the form of spatial correlations seen when the pump strength is tuned to within the window where KPZ behaviour is expected at all length scales, suggests that this regime should be easy to distinguish despite the small range of pump strengths for which it occurs.  Overall, our work strongly suggests the viability of polariton OPO in semiconductor microcavities as an experimental platform for realising and exploring KPZ physics in two dimensions, and gives an indication of how the parameters of such a system may be chosen in order to achieve this.  

\begin{acknowledgements}
We gratefully acknowledge financial support from QuantERA InterPol and EPSRC (Grant No. EP/R04399X/1 and No. EP/K003623/2).
\end{acknowledgements}

\appendix

\section{Filtering and correlations of the signal mode}\label{appendix:Filter}

To isolate the signal mode, we filter in momentum space, ultimately only considering momenta $\vect{k}$ within a square of side length \mbox{$(k_p - k_s)$} centred on the maximum of the signal mode at $\vect{k} = (k_s , 0)$, i.e.~momenta in the range $(3k_s-k_p)/2 \leq k_x \leq (k_p+k_s)/2$ and $(k_s-k_p)/2 \leq k_y \leq (k_p-k_s)/2$.  As mentioned in section \ref{section:TWA}, we label the resulting momentum-filtered signal field in \emph{real} space as $\Psi_s\!\left(\vect{x},t\right)$, with $g^{(1)}_s\!\left(\vect{r}\right)$ being its first order spatial correlation function.  

While the first order correlation of the signal is defined by equation \eqref{g1def}, by using the properties of the Fourier transform, this can be efficiently calculated in momentum space as:
\beq
g^{(1)}_s\!\left(\vect{r}\right) = \frac{\langle\Psi^*\!\left(\vect{k},t\right)\Psi\!\left(\vect{k},t\right)e^{i\vect{k}\cdot\vect{r}}\rangle_{\vect{k}\in\mathrm{signal}} - \frac{\delta_{\vect{r}, \vect{0}}}{2}}
{\langle\Psi^*\!\left(\vect{k},t\right)\Psi\!\left(\vect{k},t\right)\rangle_{\vect{k}\in\mathrm{signal}} - \frac{1}{2}} \, , \label{g1calc}
\eeq
where $\Psi\!\left(\vect{k},t\right)$ represents the 2D Fourier transform of the stochastic complex number field $\Psi\!\left(\vect{x},t\right)$, and $\langle...\rangle_{\vect{k}\in\mathrm{signal}}$ represents averaging over both stochastic realisations and all momenta $\vect{k}$ within the signal filter as defined above.

\section{Coefficients of the KPZ equation}\label{appendix:Coef}

Here we explore the coefficients and length scales associated with the KPZ equation \eqref{aKPZ}.  In table \ref{KPZcoeff}, we list the coefficients of the KPZ equation, along with the derived quantity $g_{KPZ}$, which we obtain for our microscopic model using the method described in \cite{PhysRevX.7.041006}, corresponding to the parameters used in our numerical solution of the full microscopic model, for each of the pump strengths we have investigated.  Again, these analytical calculations also require the signal momentum, which we take from the simulation results in each case.  Note that while the value of $g_{KPZ}$ for $f_p = 0.026$ (marked with an *) is actually quite large, we do not expect to observe behaviour corresponding to KPZ in this case as the KPZ equation is unstable here due to the negative values of the diffusion coefficients $D_x, D_y$.  

\begin{table}[htb]
\begin{center}
\begin{tabular}{|c|c|c|c|c|c|c|}
\hline
$f_p$ & $D_x$ & $D_y$ & $\lambda_x$ & $\lambda_y$ & $\Delta$ & $g_{KPZ}$ \\
\hline
0.051 & 0.5827 & 0.6751 & -0.5430 & -0.4805 & 0.0394 & 0.0546 \\
0.0327 & 0.0337 & 0.0469 & -0.5247 & -0.4683 & $6.4509 \times 10^{-4}$ & 2.9717 \\
0.026 & -0.0068 & -0.0111 & -0.4466 & -0.4284 & $9.3462 \times 10^{-4}$ & 464* \\
\hline
\end{tabular}
\end{center}
\caption{Pump strengths $f_p$ compared in section \ref{section:Compar}, along with the corresponding analytically calculated values of the coefficients of the KPZ equation at the signal momentum found in numerical simulations, and the quantity $g_{KPZ}$ derived from them.   \label{KPZcoeff}}
\end{table}

Derived in previous work on KPZ in incoherently driven polariton systems \cite{PhysRevX.5.011017,PhysRevB.94.104520}, in the isotropic case of $D_x = D_y = D$ and $\lambda_x = \lambda_y = \lambda$, the approximate length scales $L_v$ and $L_*$ at which the KPZ vortex unbinding phase and KPZ scaling phase (without vortices) are best estimated to appear, respectively, are given by
\beq
L_v = a_v e^{\left|\frac{2D}{\lambda}\right|}, \quad L_* = a_*e^{\frac{8\pi}{g_{KPZ}}}, \label{lengths}
\eeq
where $a_v$ and $a_*$ are corresponding microscopic length scales.  If $L_v > L_*$ then the KPZ scaling phase is expected to be visible for intermediate system sizes \mbox{$L_*<L<L_v$}, but if $L_v < L_*$ then the vortex unbinding should destroy the KPZ scaling phase or any other quasi-ordered phase for all $L > L_v$ \cite{PhysRevB.94.104520}.  In the incoherently driven system, $L_v$ cannot be made meaningfully larger than $L_*$ at drive strengths above the BKT threshold, and for the typical parameters of real microcavities both length scales are much larger than realistic system sizes \cite{PhysRevX.5.011017,PhysRevB.94.104520}, hence why here and in \cite{PhysRevX.7.041006} we instead investigate the OPO regime.  For the KPZ parameters of $f_p = 0.0327$ in table \ref{KPZcoeff}, using the values of $D_x$ and $\lambda_x$ respectively for $D$ and $\lambda$ in \eqref{lengths}, gives the corresponding length scales as $L_v \sim a_v$ and $L_* \sim 10^3 a_*$, which would imply the vortex dominated phase should win out unless $a_v \gg a_*$.  This leads to an important question, as to whether the steady state without vortices we observe in our main result, starting from coherent initial conditions, is the true steady state for this system size, or just an extremely long lived metastable state.  It should be noted that the previous analytical work \cite{PhysRevX.7.041006} was unable to consider the affect of vortices beyond simply looking at the approximate length scales \eqref{lengths}.  

In Fig.~\ref{vortexdecay}, we see that even when starting from highly disordered initial conditions, the system still evolves towards the vortex free steady state at long times.  This suggests that, either through affecting the microscopic length scales $a_v,a_*$ or otherwise, the full behaviour of polariton OPO captured by our simulations does indeed result in the KPZ scaling phase without vortices being the true steady state.  There are two possible contributing factors to the observed behaviour which we can identify.  The first is the small value of the diffusion coefficients $D_x,D_y$.  It can be seen from table \ref{KPZcoeff}, that this is the major contributing factor to the larger $g_{KPZ}$, and hence smaller $L_*$, within the KPZ window, and also shrinks $L_*$ much faster than it shrinks $L_v$.  While for those values at $f_p = 0.0327$ this is not enough to give $L_* < L_v$ from the approximate formulae \eqref{lengths}, assuming $a_v \sim a_*$, it may cause $L_* < L_v$ if $a_v$ is large or the true value of $L_v$ is otherwise underestimated by \eqref{lengths}.  A second factor that may suppress the vortex dominated phase is the anisotropy of the system.  While the anisotropy of the KPZ coefficients in table \ref{KPZcoeff} is relatively small, the inherent anisotropy of the polariton OPO system may still result in anisotropic interactions between vortices, which could cause the overall behaviour of vortices to differ from that predicted assuming isotropic interactions.

\section{Investigating the exponent $\chi$}\label{appendix:Exp}

\begin{figure}[h!]
\includegraphics[width=\columnwidth]{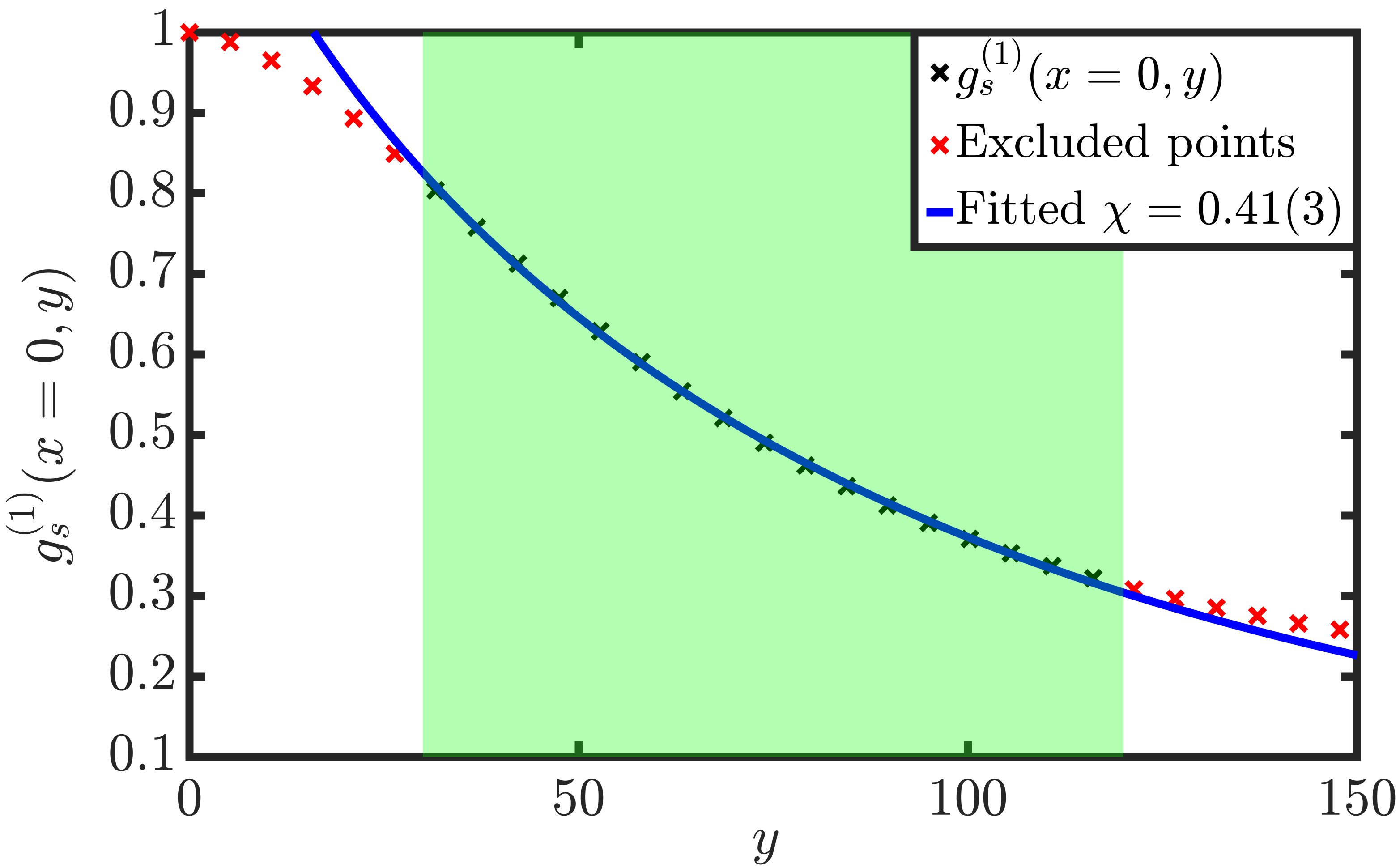}
\caption{$g^{(1)}_s\!\left(x = 0, y\right)$ with stretched exponential fit where the exponent $\chi$ is also a fitting parameter.  Fitting in the same (green shaded) region $30 < y < 120$ as used for the fits in Fig.~\ref{g1Allfits327} gives a value of $\chi = 0.41(3)$, where error in last digit represents $95\%$ confidence bounds.  
\label{freeChi}}
\end{figure}

For the fitting results presented in section \ref{section:Fitting}, two parameter fits were used to compare all the different models.  Consequently, the value of $\chi$ for the stretched exponential fit was fixed to its expected value of $\chi = 0.39$.  Here, we perform further checks that the decay of spatial correlations in the $y$ direction actually behaves in the way predicted by the corresponding KPZ equation by also fitting for the value of the characteristic exponent $\chi$.  A plot of this fit over the same fitting range used previously is shown in Fig.~\ref{freeChi}.  We find a fitted value of $\chi = 0.41(3)$, in agreement with the value $\chi \approx 0.39$ known for the 2D KPZ universality.  

\begin{figure}[h!]
\includegraphics[width=\columnwidth]{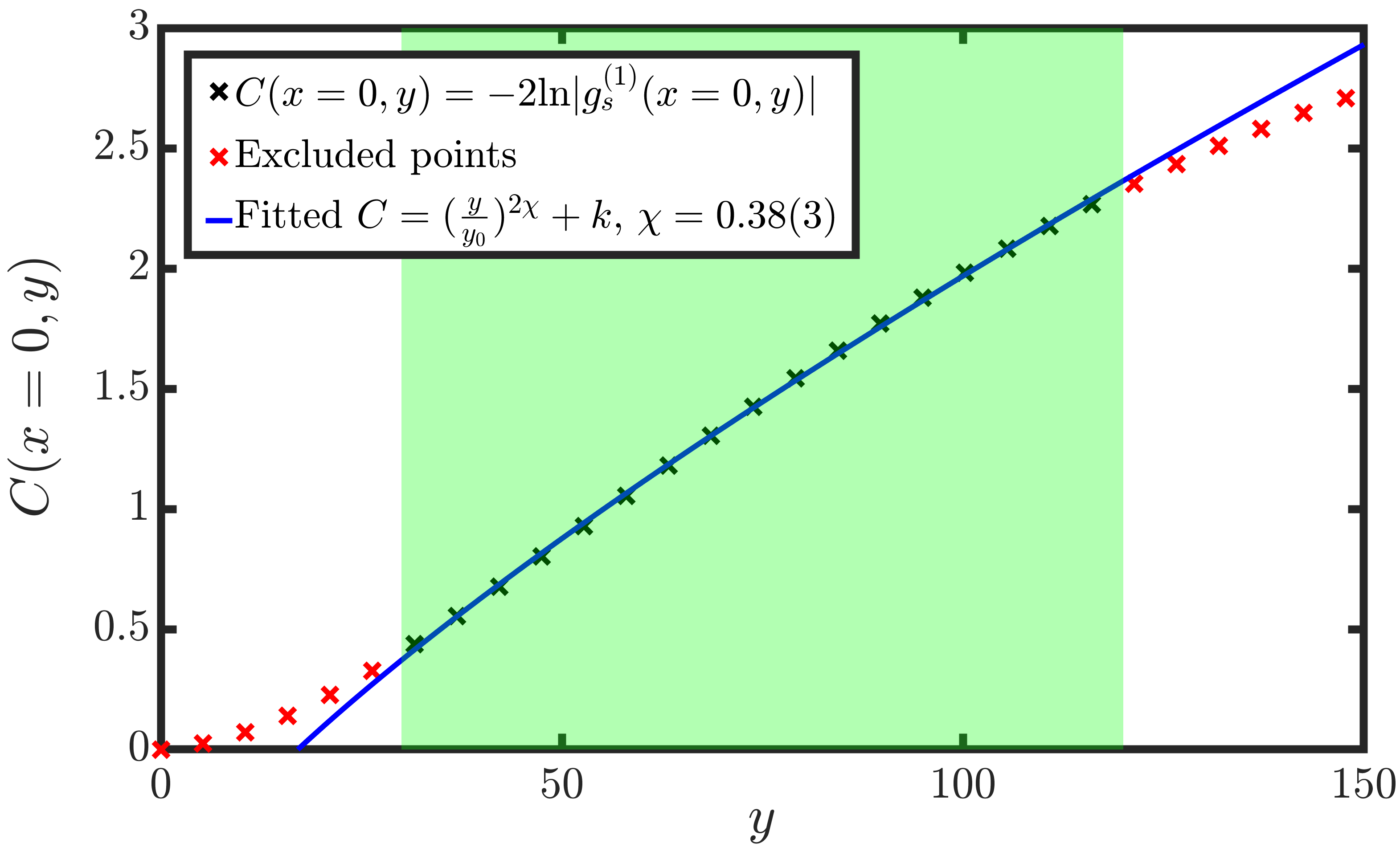}
\caption{Fit for power law scaling $C = (y/y_0)^{2\chi} + k$ of phase correlations \mbox{$C(x = 0, y) = -2\mathrm{ln}|g^{(1)}_s\!\left(x = 0, y\right)\!|$}.  Fitting in the same (green shaded) region $30 < y < 120$ as used for the fits in Fig.~\ref{g1Allfits327} gives a value of $\chi = 0.38(3)$, where error in last digit represents $95\%$ confidence bounds.  
\label{PhaseCfit}}
\end{figure}

We can also perform a similar analysis by looking at the connected correlation function of the phase $\theta(\vect{x}, t)$ of the filtered signal mode
\beq
C(\vect{r}) = \langle \left[ \theta(\vect{R}+\vect{r}, t) - \theta(\vect{R}, t) \right]^2 \rangle - \langle \theta(\vect{R}+\vect{r}, t) - \theta(\vect{R}, t) \rangle^2 \, ,
\eeq
where averages are over realisations, the position $\vect{R}$, and times $t$ within the steady state.  We can calculate this from the first order correlations of the signal as \mbox{$C(x = 0, y) = -2\mathrm{ln}|g^{(1)}_s\!\left(x = 0, y\right)\!|$} \cite{Deligiannis_2021}.  A fit of this to a power law $C = (y/y_0)^{2\chi} + k$, with $y_0$, $k$ and $\chi$ determined by the fitting, is shown in Fig.~\ref{PhaseCfit}.  This finds the value of the critical exponent as $\chi = 0.38(3)$, again in good agreement with previously determined value of that exponent for 2D KPZ.  

\section{Anisotropy in correlations within the KPZ window}\label{appendix:Xdir}

\begin{figure}[h!]
\includegraphics[width=\columnwidth]{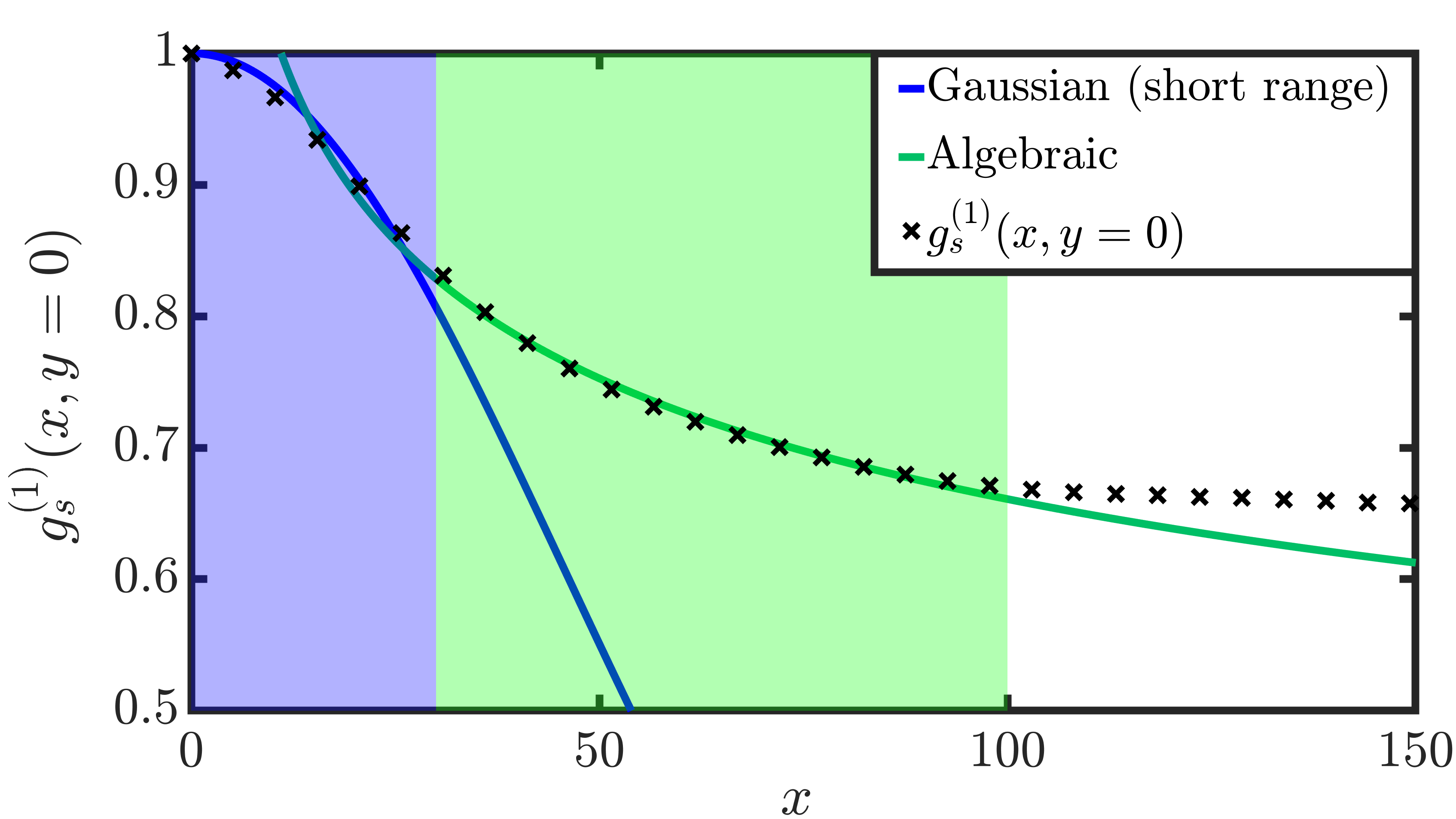}
\includegraphics[width=\columnwidth]{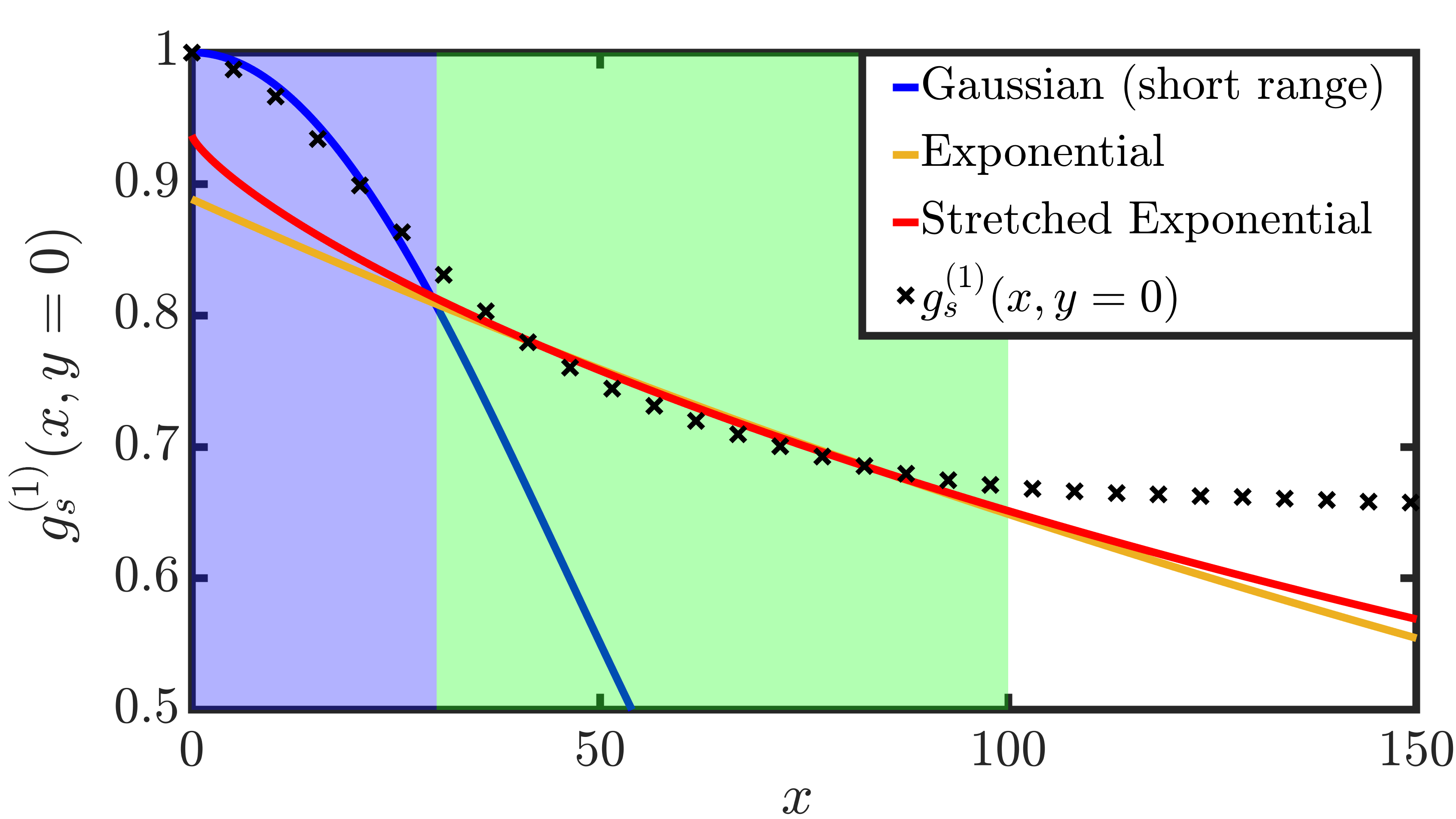}
\caption{$g^{(1)}_s\!\left(x, y = 0\right)$ with fits to algebraic (green line, top panel), exponential (yellow, bottom panel), and stretched exponential (red, bottom panel) decay.  Green shaded region indicates the points included in the fit.  Blue line gives Gaussian fit to short range correlations within blue region, used to determine the lower fitting bound, as in Fig.~\ref{g1Allfits327}.
\label{g1sxfits}}
\end{figure}

In this section, we explore in greater detail the discrepancy between the correlations in the $x$ direction at \mbox{$f_p = 0.0327$}, and the behaviour predicted by KPZ that is seen in the $y$ direction.  In Fig.~\ref{g1sxfits}, we show $g^{(1)}_s\!\left(x, y = 0\right)$ with fits to algebraic, exponential and stretched exponential decay.  It can be seen in Fig.~\ref{g1sxfits} that $g^{(1)}_s\!\left(x, y = 0\right)$ saturates quickly to a value of approximately 0.66; as a result we reduce the fitting region to $30 < x < 100$, to exclude more of the furthest points compared to that used for $g^{(1)}_s\!\left(x =0, y\right)$ in section \ref{section:Fitting}.  By the coefficient of determination $R^2$, the best fit is the algebraic ($R^2 = 0.9938$); exponential and stretched exponential fits have $R^2 = 0.9448$ and $R^2 = 0.9592$ respectively.  In all cases these are weaker than the best fits for $g^{(1)}_s\!\left(x =0, y\right)$.  

\begin{figure}[h!]
\includegraphics[width=\columnwidth]{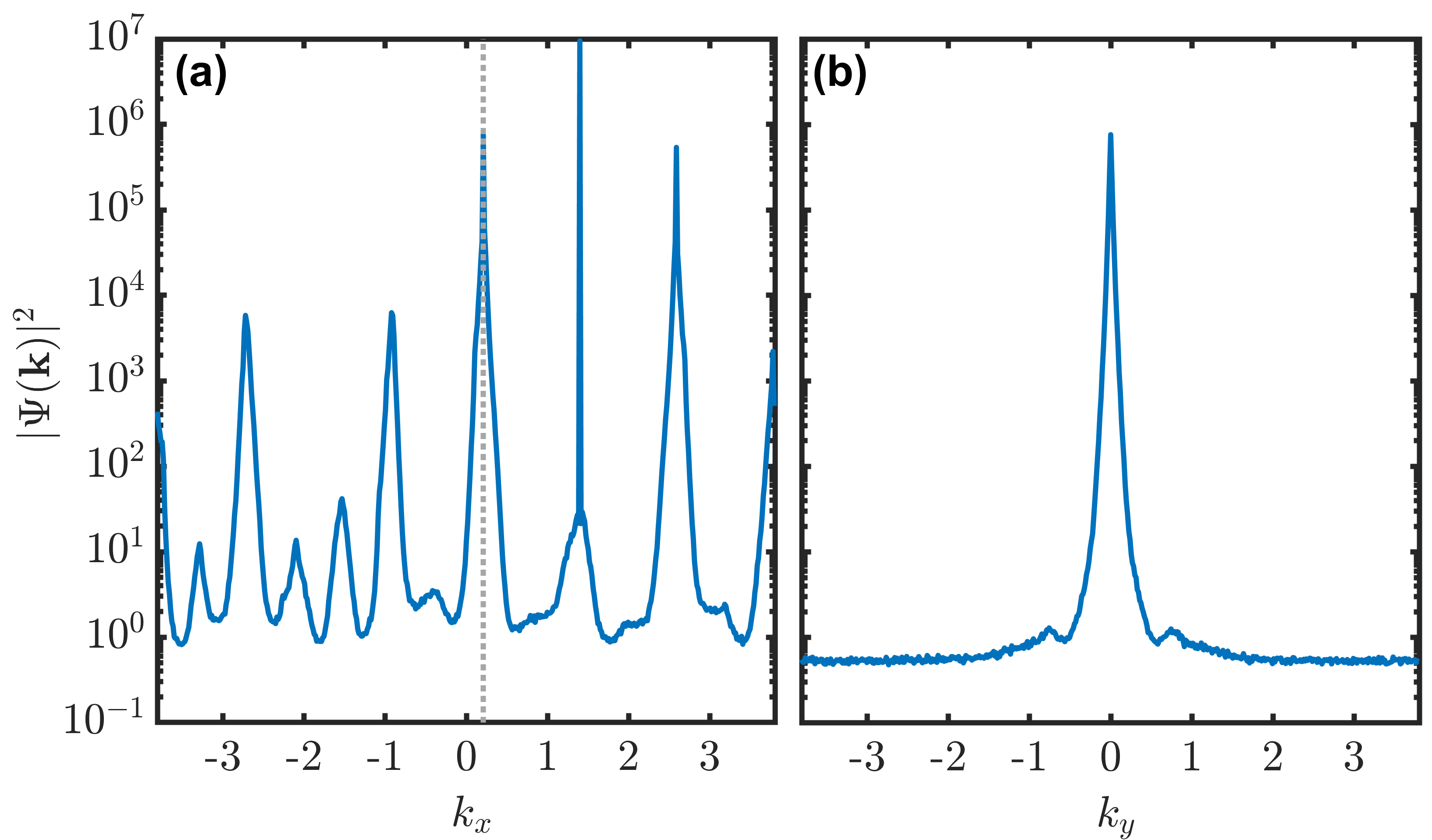}
\caption{Steady state momentum distribution of OPO $|\Psi(\vect{k})|^2$ (log scale) at \mbox{$f_p = 0.0327$}.  (a) Distribution in $k_x$ along \mbox{$k_y = 0$}.  Vertical dotted line shows signal momentum \mbox{$k_x = k_s = 0.2084$}.  (b) Distribution in $k_y$ at $k_x = k_s = 0.2084$.  
\label{kprofs}}
\end{figure}

In section \ref{section:Fitting}, we state that we suspect the reason for this unexpected anisotropy is due to the presence of satellite states generated by secondary scattering of the signal mode along $k_x$.  Here we will attempt to make this argument more concrete.  Firstly, examining table \ref{KPZcoeff} once again, we can see that the coefficients of the KPZ equation are only very weakly anisotropic.  This suggests that the source of the anisotropy in the correlations lies beyond the approximations under which the OPO maps to the KPZ equation.  In Fig.~\ref{kprofs}, we show the momentum distribution at $f_p = 0.0327$.  It can be seen in Fig.~\ref{kprofs}a that this case has strong satellite states distributed along $k_x$.  We argue that the reason for the unexpected  behaviour of the correlations in the $x$ direction might be due to the presence of the satellite states interfering with the KPZ phase dynamics that would be expected in their absence.  Although  the satellite states are excluded from the calculation of $g^{(1)}_s\!\left(x, y = 0\right)$ by the filter in momentum space which we use to isolate the signal mode, it seems that their presence might still have a strong effect on the behaviour of the signal's phase.  In Fig.~\ref{kprofs}b, we can see that there are no such complications to the structure of the signal mode along $k_y$, hence why the stretched exponential decay can be observed in the $y$ direction unhindered.  We believe this indicates that we are observing the KPZ phase despite the more complicated form of correlations in the $x$ direction.

\section{Convergence checks}\label{appendix:Conv}

To ensure the validity of our findings, we check convergence of the results in time (convergence to steady state), the number of stochastic realisations used, and system size.  

\begin{figure}[h!]
\includegraphics[width=\columnwidth]{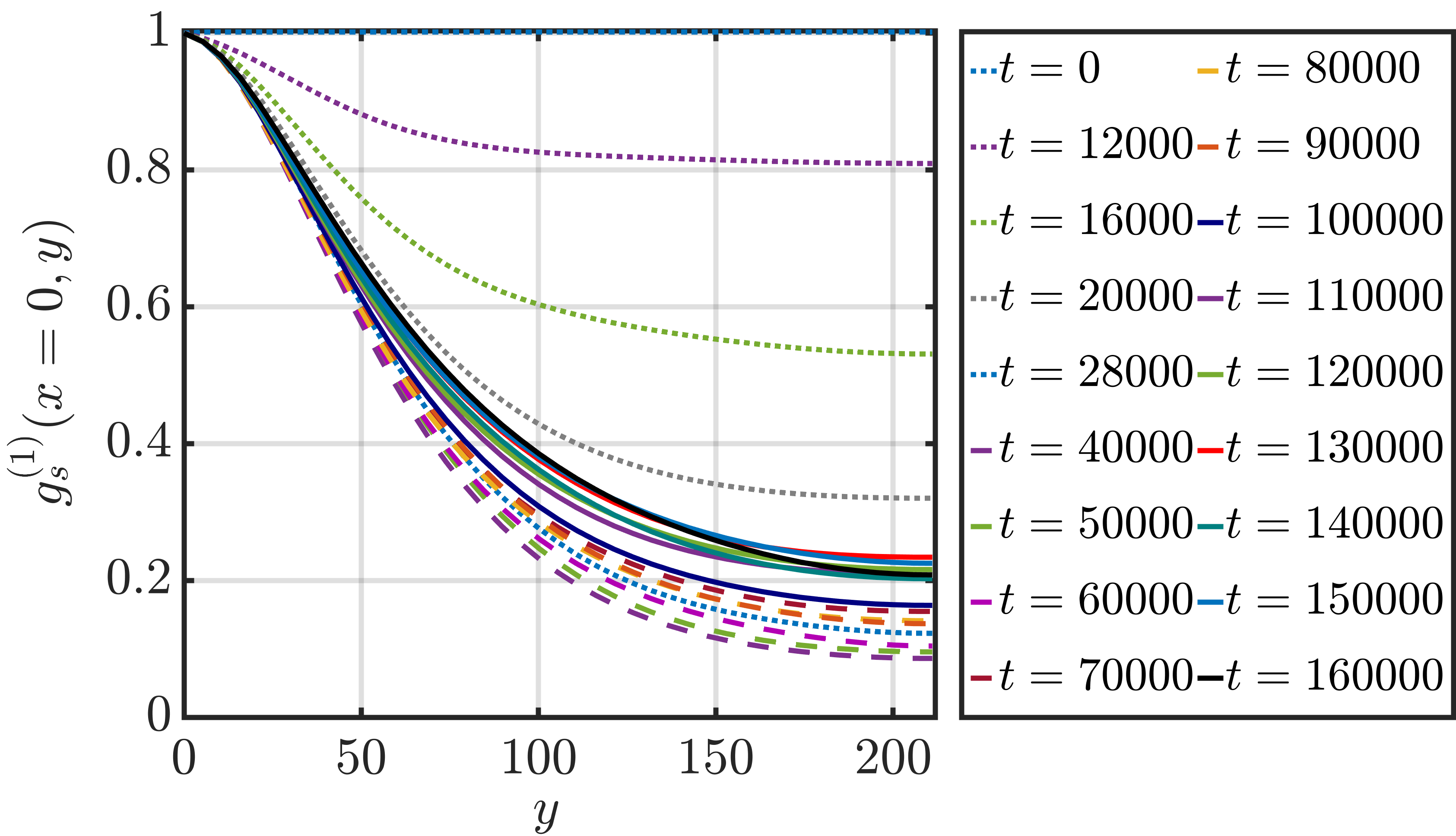}
\caption{Calculated $g^{(1)}_s\!\left(x = 0, y\right)$ at different times from $t = 0$ to $t = 160000$.
\label{tconv}}
\end{figure}

\subsection{Convergence to steady state}

We use the mean-field steady-state as the initial condition for our stochastic dynamics.  Our prior analysis of stochastic simulations for the OPO system \cite{PhysRevX.5.041028} close to the BKT transition showed that the steady-state does not depend on the initial conditions.   Different observables can take different times to reach a steady state, with $g^{(1)}_s\!\left(\vect{r}\right)$ being one of the slowest to converge.  \mbox{Fig.~\ref{tconv}} shows the evolution of $g^{(1)}_s\!\left(x = 0, y\right)$ in time.  Beyond around $t =120000$, $g^{(1)}_s\!\left(x = 0, y\right)$ stops drifting and remains stable except for small fluctuations.  All other results are therefore obtained by averaging over the steady state from $t =120000$ to $t =160000$.

\subsection{Convergence with number of realisations}

\begin{figure}[h!]
\includegraphics[width=\columnwidth]{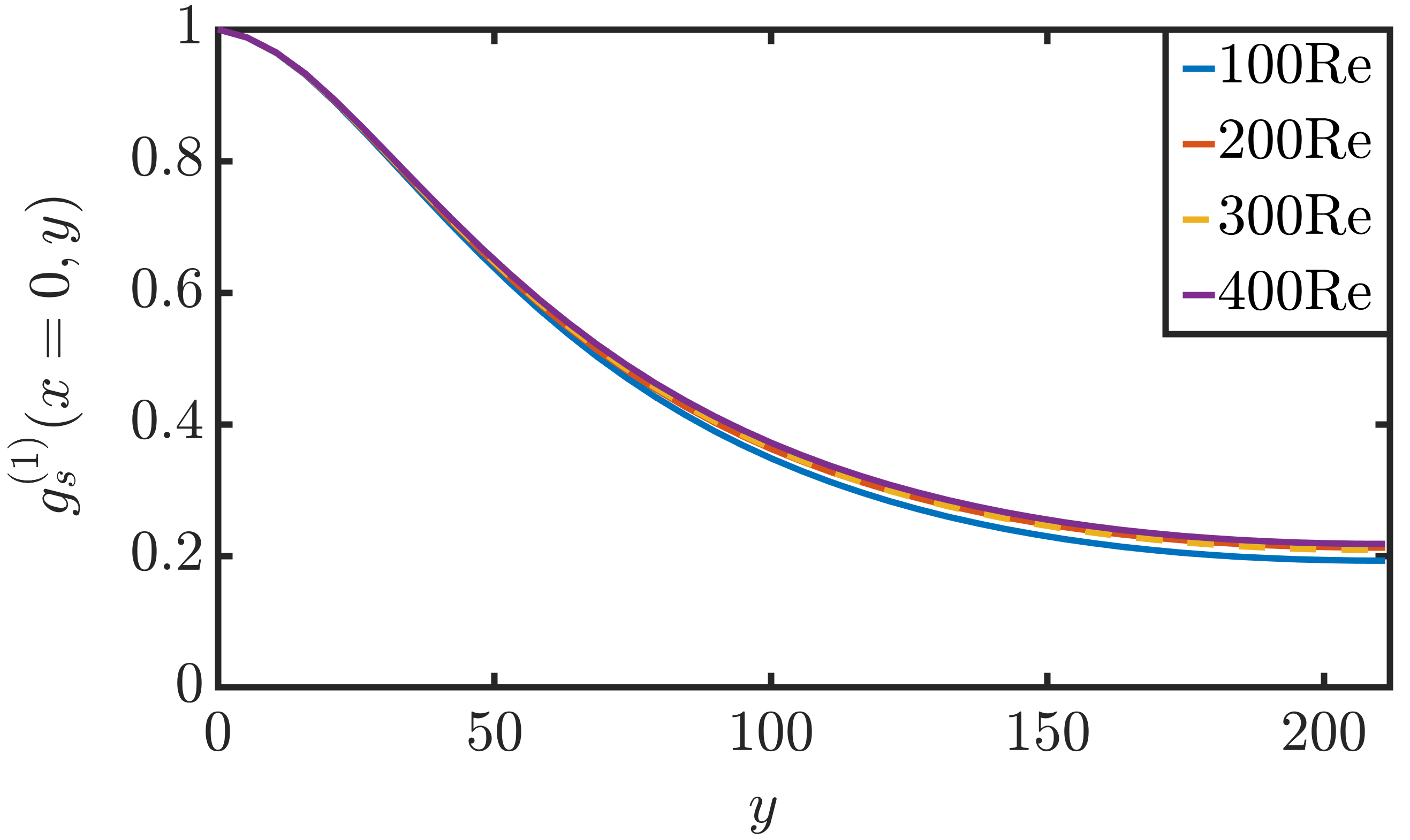}
\caption{Calculated $g^{(1)}_s\!\left(x = 0, y\right)$ for different numbers of realisations used between 100 and 400.  Each case is also time averaged over the steady state.  
\label{ReComp}}
\end{figure}

Since results from the TWA method are produced by averaging over stochastic realisations, it is important to check that we have used enough realisations to sufficiently sample the underlying distribution, and hence give results that do not depend on the exact number of realisation used.  Different physical quantities require different numbers of realisations to converge; for example, the momentum distributions $|\Psi\!\left(\vect{k},t\right)|^2$ can often show minimal differences between individual realisations, but the correlation function $g^{(1)}_s\!\left(\vect{r}\right)$ typically requires a large number of realisations to fully converge.  We used a total of 400 realisations for the main result at $f_p = 0.0327$.  To check that this is sufficient we compare the form of 
$g^{(1)}_s\!\left(\vect{r}\right)$ when calculated with specific numbers of realisations (see Fig.~\ref{ReComp}).  We see that beyond 200 realisations the form of $g^{(1)}_s\!\left(\vect{r}\right)$ does not change significantly, suggesting that the 400 realisations is sufficient for capturing the behaviour of correlations.  

\subsection{Convergence with system size}

To check that our results do not depend on the system size, we run the simulations again for a slightly smaller system with $N = 384$, $L = 165.85464$.  Note that the specific values of $N$ and $L$ are chosen such as to make sure that the signal momentum 
$k_s = 0.2084$, chosen by the larger system, still lies on the numerical grid in momentum space for the smaller system, as the calculated KPZ non-linearity shown in Fig.~\ref{fig_KPZnl} is generally dependent on the exact value of $k_s$, and so we must allow for it to remain the same to truly compare different system sizes.  

\begin{figure}[h!]
\includegraphics[width=\columnwidth]{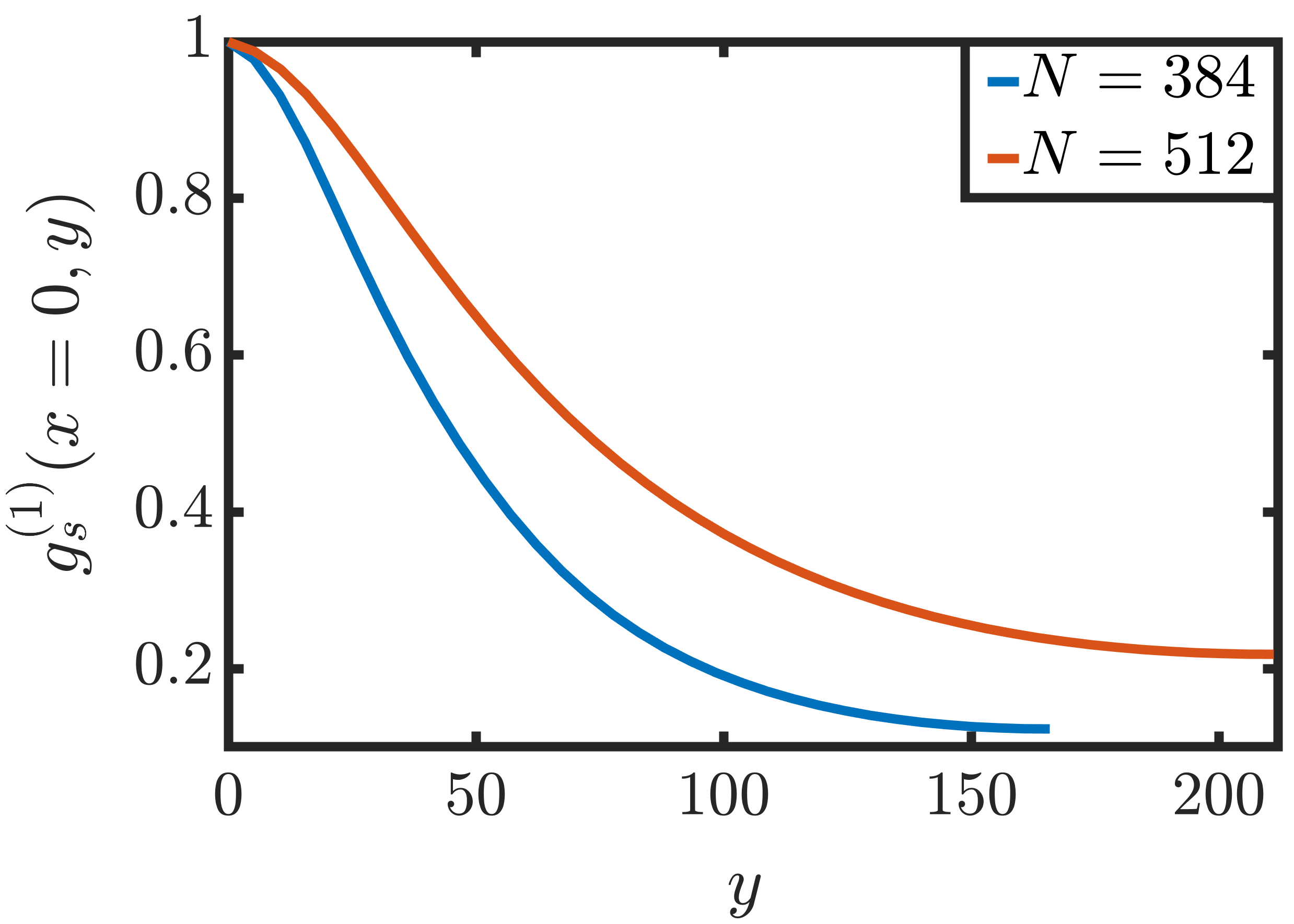}
\caption{Calculated $g^{(1)}_s\!\left(x = 0, y\right)$ for different system sizes $N = 512$ ($L = 211.08772$, red line) and $N = 384$ ($L = 165.85464$, blue line).  Each case is also time averaged over the steady state.  
\label{SizeComp}}
\end{figure}

\begin{figure}[h!]
\includegraphics[width=\columnwidth]{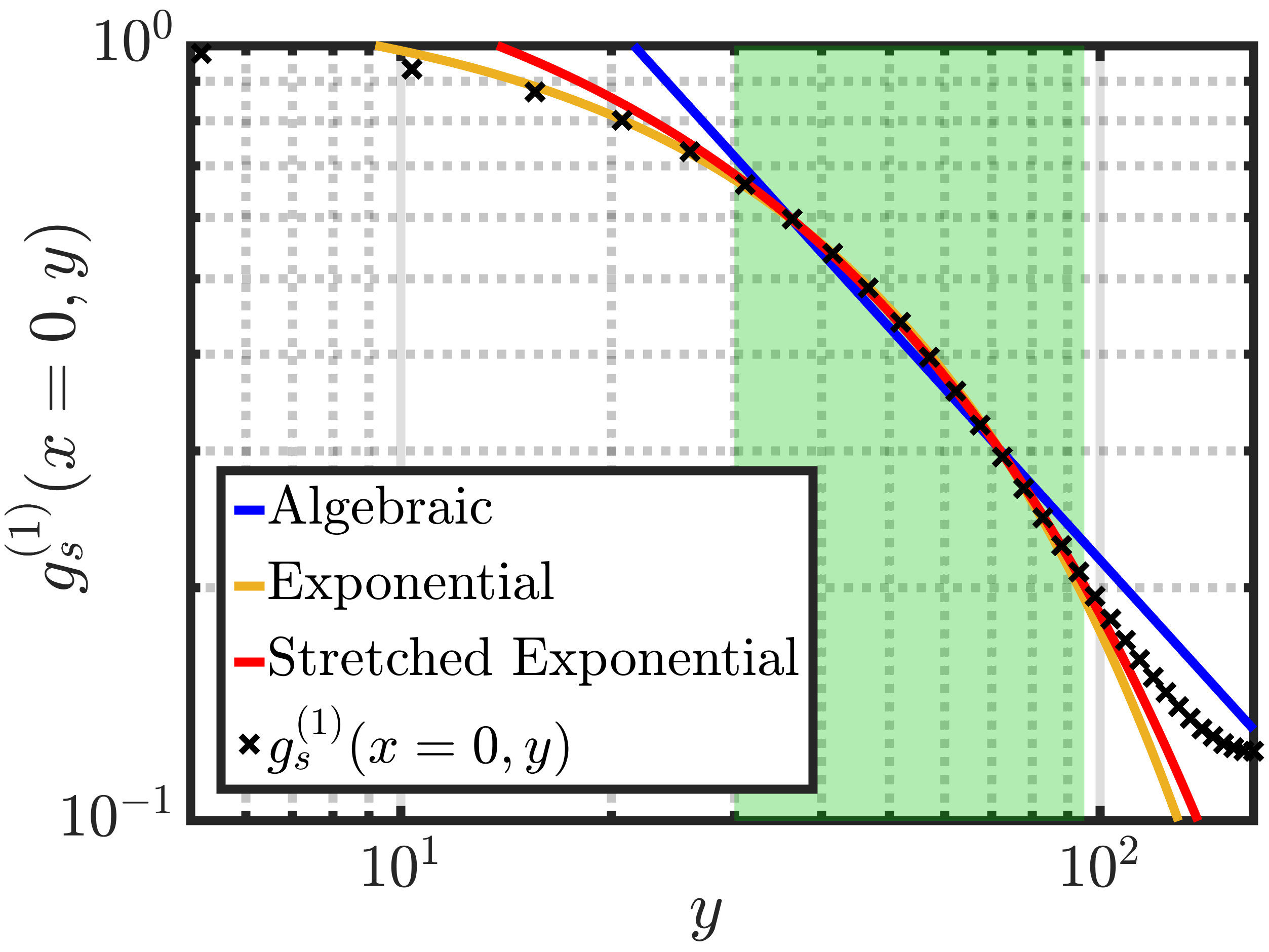}
\caption{$g^{(1)}_s\!\left(x = 0, y\right)$ with fits to algebraic (blue line), exponential (yellow line), and stretched exponential (red line) decay, for smaller system size ($N = 384$).  Green shaded region indicates the points included in the fit.  
\label{g1smallfits327}}
\end{figure}

Although $g^{(1)}_s\!\left(x = 0, y\right)$ for the different system sizes differs in magnitude (see Fig.~\ref{SizeComp}), as can be seen from Fig.~\ref{g1smallfits327} (for $N = 384$) and Fig.~\ref{g1Allfits327} (for $N = 512$), both system sizes are independently seen to have a good fit to the stretched exponential decay with exponent $\chi = 0.39$, as predicted from the KPZ equation.  This suggests that while the exact form of the OPO produced may still be too finely dependent on the geometry of the system for the sizes we consider, the observability of the signatures of KPZ fluctuations is much less dependent on the system size.  

\begin{figure}[h!]
\includegraphics[width=\columnwidth]{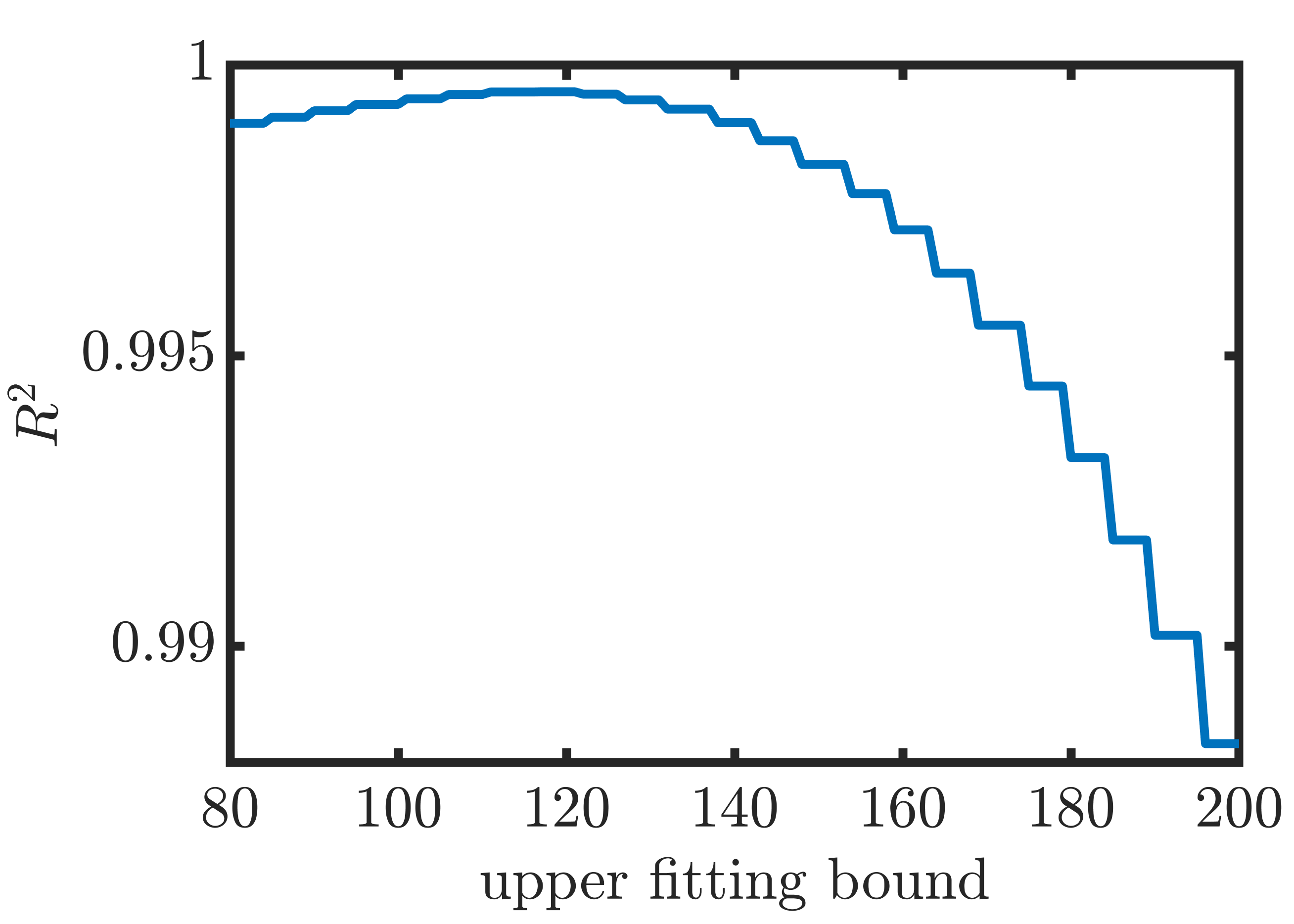}
\caption{Coefficient of determination $R^2$ for stretched exponential fit as a function of the chosen upper fitting bound.  Lower fitting bound is fixed as $y > 30$.
\label{ubcheck}}
\end{figure}

\section{Choice of fitting bounds}\label{appendix:Bounds}

Here we look at how we choose the exclusion bounds for the points included in our fitting, and how it affects the results of the stretched exponential fit.  These exclusions are needed to account for microscopic and boundary effects.  Firstly, a lower bound to the fitted region is necessary because the mapping from polariton OPO to the KPZ equation is only valid in the long range limit.  The form of the correlations at short distances is not universal and may depend on the microscopic details of the specific system.  The upper fitting bound instead solves a problem of a more practical origin: the finite size of the simulated system, and the boundary conditions imposed at the edges of that finite system.  For the simulations we use periodic boundary conditions, which tend to enhance the correlations near the edges.  

As shown in Fig.~\ref{g1Allfits327}, we can justify our choice of lower bound by fitting the expected Gaussian form of the short range correlations.  We see that $g^{(1)}_s\!\left(x = 0, y\right)$ fits well to a Gaussian form for $y < 30$, and so choose to exclude points with $y < 30$ from our fits for the long range form of the correlations.  The appropriate upper fitting bound is a bit harder to determine robustly. In practice, we should exclude the area which is affected by the periodic boundary conditions i.e.~where the correlations start to grow due to the proximity of the next unit cell.  However, as we can see in Fig.~\ref{ubcheck}, the quality of our stretched exponential fit, as given by the coefficient of determination $R^2$ does not depend strongly on our choice of upper bound until around $y = 140$, beyond which it begins to fall much more rapidly.  We therefore feel comfortable choosing to exclude points with $y > 120$ from our stretched exponential fit (which maximises $R^2$) to remove the edge effects, knowing that while the choice is somewhat arbitrary, it does not significantly affect our results.


%

\end{document}